\documentclass[amsmath,amssymb,reprint,groupedaddress,showpacs,aps,prx]{revtex4-1}
\usepackage{graphicx}
\usepackage{xcolor}
\usepackage{soul}
\sethlcolor{yellow}
\usepackage{bbold }
\usepackage{caption}
\usepackage{subcaption}
\usepackage{dsfont }
\definecolor{red}{rgb}{1,0.,0}

\usepackage{hyperref}

\begin{document}

\title{Continuous-variable quantum key distribution field-test with true local oscillator}

\author{Brian P. Williams*}
\affiliation{Quantum Information Science Section, Oak Ridge National Laboratory, Oak Ridge, Tennessee USA 37831}
\email{williamsbp@ornl.gov}
\author{Bing Qi}
\affiliation{Quantum Information Science Section, Oak Ridge National Laboratory, Oak Ridge, Tennessee USA 37831}
\email{qib1@ornl.gov}
\author{Muneer Alshowkan}
\affiliation{Quantum Information Science Section, Oak Ridge National Laboratory, Oak Ridge, Tennessee USA 37831}
\email{alshowkanm@ornl.gov}
\author{Philip G. Evans}
\affiliation{Quantum Information Science Section, Oak Ridge National Laboratory, Oak Ridge, Tennessee USA 37831}
\email{evanspg@ornl.gov}
\author{Nicholas A. Peters}
\affiliation{Quantum Information Science Section, Oak Ridge National Laboratory, Oak Ridge, Tennessee USA 37831}
\email{petersna@ornl.gov}

\begin{abstract}
%
Continuous-variable quantum key distribution (CV-QKD) using a true local (located at the receiver) oscillator (LO) has been proposed to remove any possibility of side-channel attacks associated with transmission of the LO as well as reduce the cross-pulse contamination. Here we report an implementation of true LO CV-QKD  using ``off-the-shelf'' components and conduct QKD experiments using the fiber optical network at Oak Ridge National Laboratory. A phase reference and quantum signal are time multiplexed and then wavelength division multiplexed with the classical communications which ``coexist'' with each other on a single optical network fiber. This is the first demonstration of CV-QKD with a receiver-based true LO over a deployed fiber network, a crucial step for its application in real-world situations.

\end{abstract}
\maketitle

\section{Introduction}\label{intro}

Continuous variable quantum key distribution protocols based on optical coherent detection have drawn great attention for their compatibility with conventional fiber optical networks \cite{diamanti2015distributing,laudenbach2018continuous}. The well-established Gaussian modulated coherent state (GMCS) protocol \cite{grosshans2003quantum} can be realized with off-the-shelf telecom components and has been recently demonstrated over 200~km of low-loss fiber \cite{zhang2020long}. With the emergence of CV-QKD based upon discrete modulation \cite{zhao2009asymptotic, leverrier2009unconditional, bradler2018security, ghorai2019asymptotic, lin2019asymptotic, kaur2021asymptotic}, it is conceivable that both classical communications and QKD could be implemented with a single communications system, either time-multiplexed or through simultaneous transmission~\cite{qi2016simultaneous}.

Several desirable features of CV-QKD can be attributed to its homodyne detection scheme which uses a strong local oscillator (LO) to interfere a weak quantum signal at a beam splitter before photodetection with a shot-noise limited optical detector. On one hand, the LO acts as an ``amplifier'' and the resulting interference signal is strong enough to be detected by conventional photodiodes working at room temperature. On the other hand, the LO also functions as a highly selective ``filter'' and can effectively suppress out-of-band noise photons. This intrinsic filtering makes CV-QKD an appealing solution for coexistence with classical communications signals through shared fiber infrastructure \cite{qi2010feasibility, kumar2015coexistence, karinou2018toward,eriksson2019wavelength}.

A major technical challenge in CV-QKD detection is carrier phase recovery: to reliably decode the information encoded on field quadratures, the receiver (Bob) needs to share a common optical phase reference with the sender (Alice). The conventional approach is to generate both the weak quantum signal and the strong LO from the same laser at Alice and send both through the same quantum channel to Bob \cite{lodewyck2007quantum, qi2007experimental, jouguet2013experimental}. More recently, a truly ``local'' LO scheme has been developed for CV-QKD \cite{qi2015generating, soh2015self}, where Bob (the receiver) generates the LO in his secure enclave from an independent laser and the carrier phase recovery is achieved by transmitting weak phase reference pulses from Alice to Bob. The true LO scheme not only removes potential side-channel vulnerabilities associated with sending the LO through the insecure quantum channel \cite{ma2013wavelength, huang2013quantum, jouguet2013preventing}, but also eliminates the high-isolation multiplexing and demultiplexing techniques required to isolate the strong LO from the much weaker quantum signal at the receiver to prevent cross talk. While CV-QKD with a true LO scheme has been demonstrated by several groups \cite{qi2015generating, soh2015self, huang2015high, kleis2017continuous, laudenbach2019pilot,wang2021sub}, to our knowledge, it has never been demonstrated over deployed fiber between two distant users due to the many technical challenges involved.

In this paper, we develop a CV-QKD GMCS system with a true LO scheme and evaluate its performance by transmitting quantum signals through deployed (mostly aerial) fiber between two sites at Oak Ridge National Laboratory, along with the bidirectional classical communications signals. To reduce the phase noise, we adopt a design proposed in \cite{marie2017self} where the phase reference pulse and the quantum signal are split from a common laser pulse. This design removes the dependence of phase noise on the linewidth of the laser and has been demonstrated experimentally \cite{wang2018high, qi2018noise}. In our demonstration Alice and Bob are physically separated, requiring us to solve practical challenges such as timing synchronization, polarization correction, detector bias-correction, and signal-reference relative phase correction.

The paper is organized as follows: In Section II, we review the true LO protocol, in Section III we present our experimental implementation, and we detail the novelty of our true LO approach in Section IV. Finally, we present our results and discussion in Section V.  

\section{True LO CV-QKD Protocol and Challenges
\label{Local LO CV-QKD}}

\begin{figure}[]
\includegraphics[width=\linewidth]{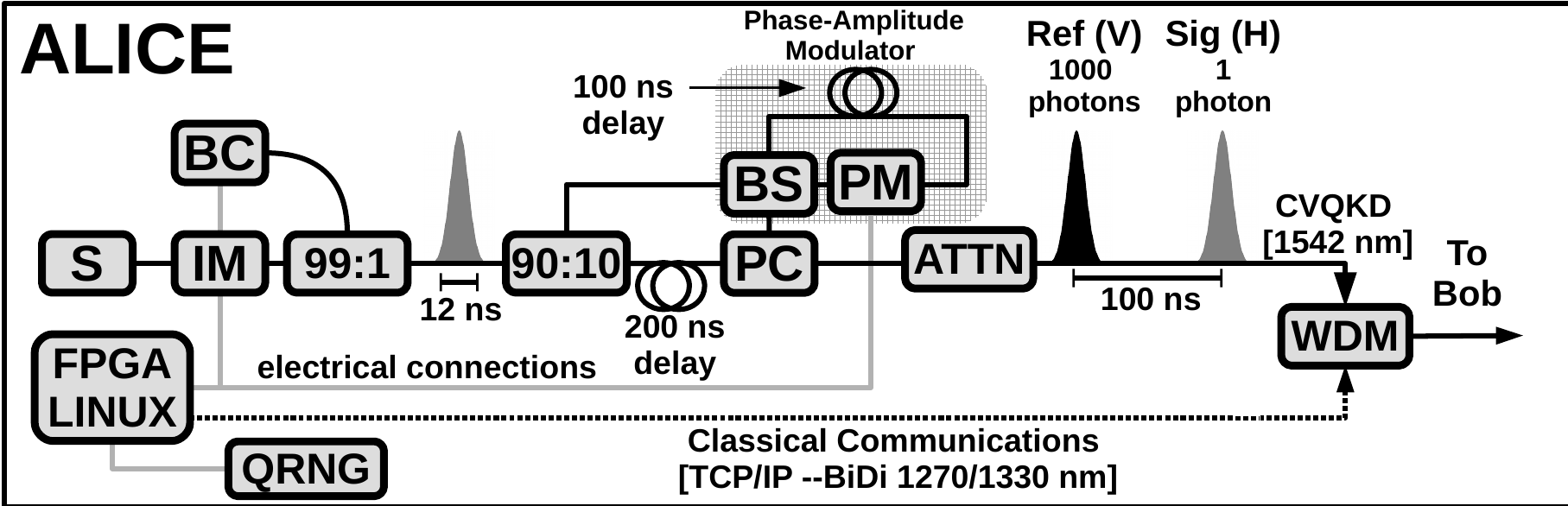}
\caption{Alice's CV-QKD transmitter consists of a source laser, optical modulators, fiber splitters/combiners, and attenuators to generate her quantum signal and reference pulse. She combines her quantum signal at 1542 nm with light for classical communications at 1270/1330 nm. See Section \ref{experiment} for a complete description.}
\label{AliceExpt}
\includegraphics[width=\linewidth]{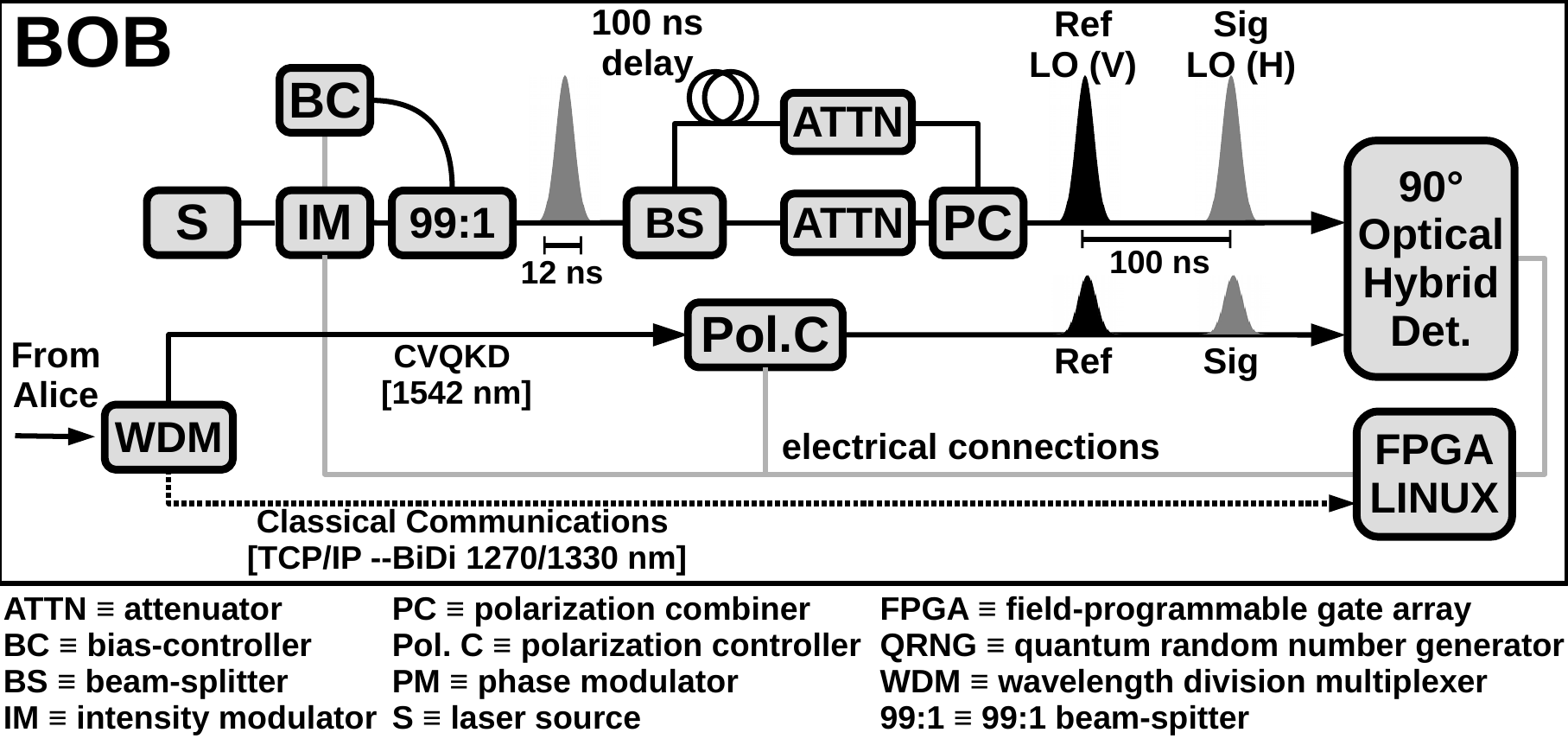}
\caption{Bob's CV-QKD receiver is similar to Alice's transmitter with the addition of a 90$^\circ$ optical hybrid detector to measure the signal and reference pulses. Bob has his own laser source, this is the true local oscillator (LO). See Section \ref{experiment} for a complete description.}
\label{BobExpt}
\end{figure}

A theoretical presentation of the true LO concept, as well as a proof-of-principle demonstration, has previously been published by Qi et al.~\cite{qi2015generating}. Here, we review the true LO data recovery scheme appropriate to our practical experimental implementation illustrated in Fig. \ref{AliceExpt} and \ref{BobExpt}. 

\begin{figure}[tb]
    \centering
    \includegraphics[width=\linewidth]{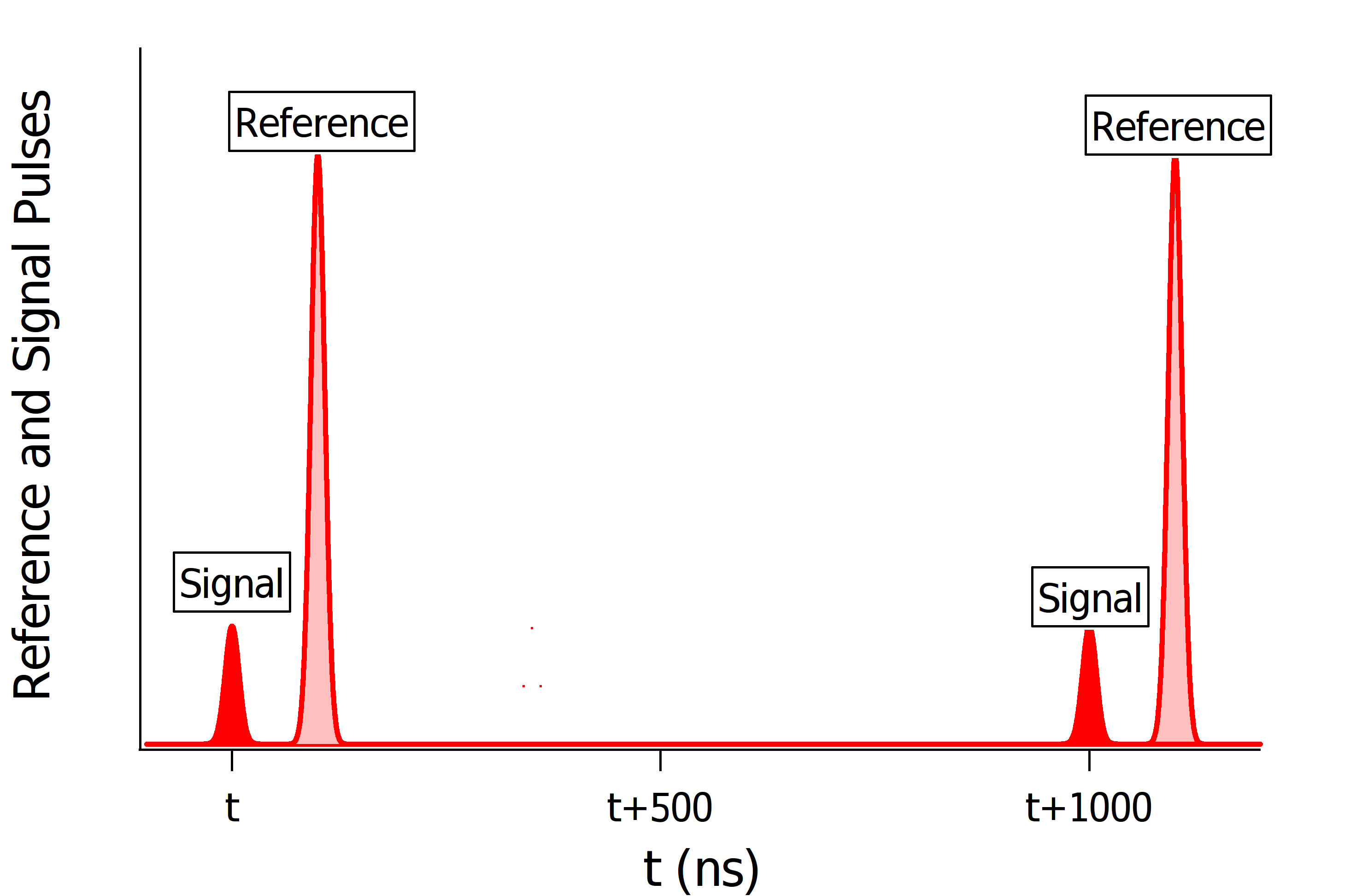}
    \caption{CV-QKD transmissions from Alice: a signal pulse (with average photon number $n \approx$ 1) at $t$ = 0 ns precedes a reference pulse (with $n \approx$ 1000) at $t$ = 100 ns. The sequence repeats after 1000 ns, providing a 1 MHz system repetition rate. The final symbol rate is 50 kSymbols/s due to post-processing overhead.}
    \label{alicePulses}
\end{figure}

\begin{figure}[tb]
\includegraphics[width=0.48\linewidth]{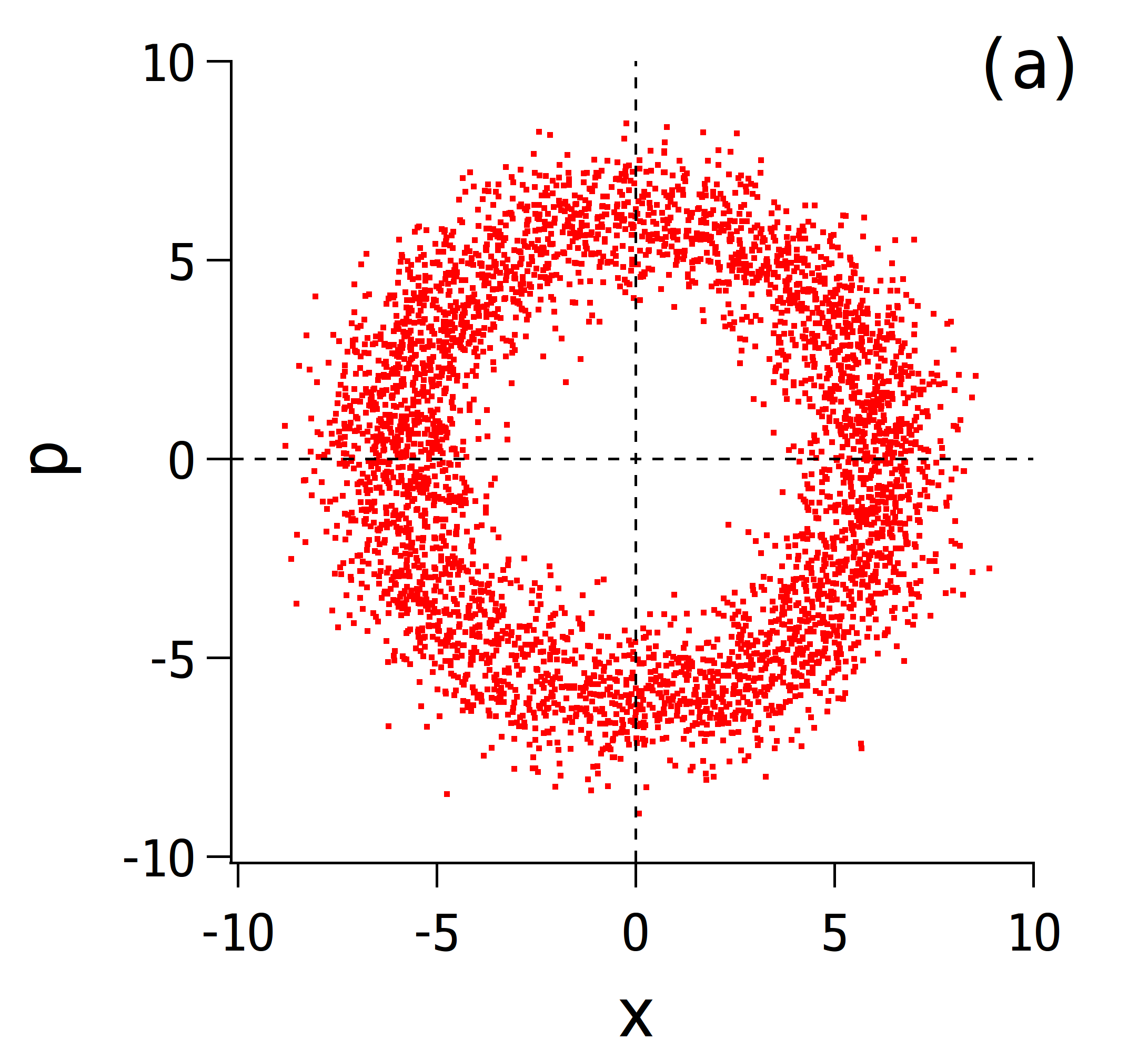}
\includegraphics[width=0.48\linewidth]{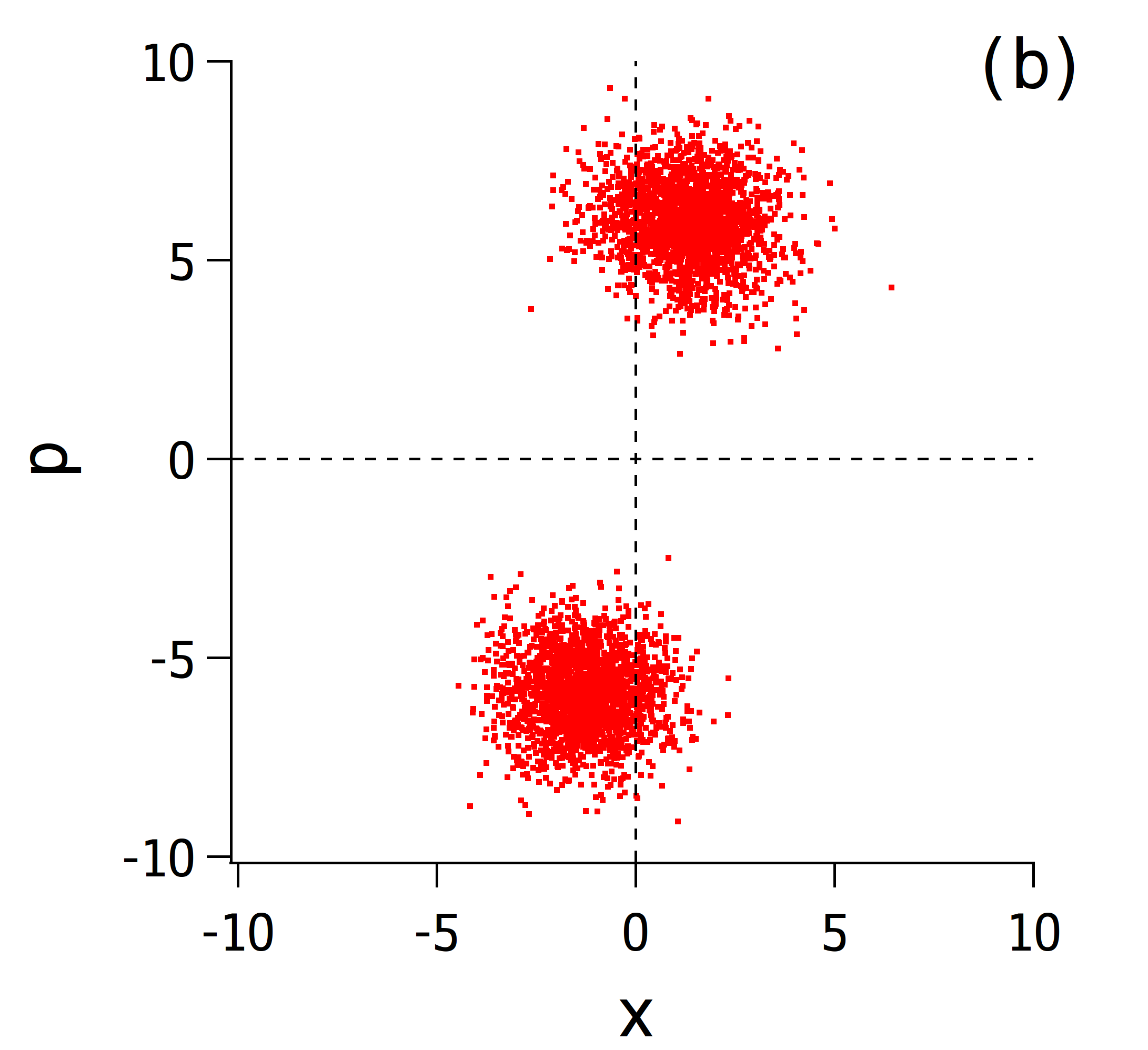}
\includegraphics[width=0.48\linewidth]{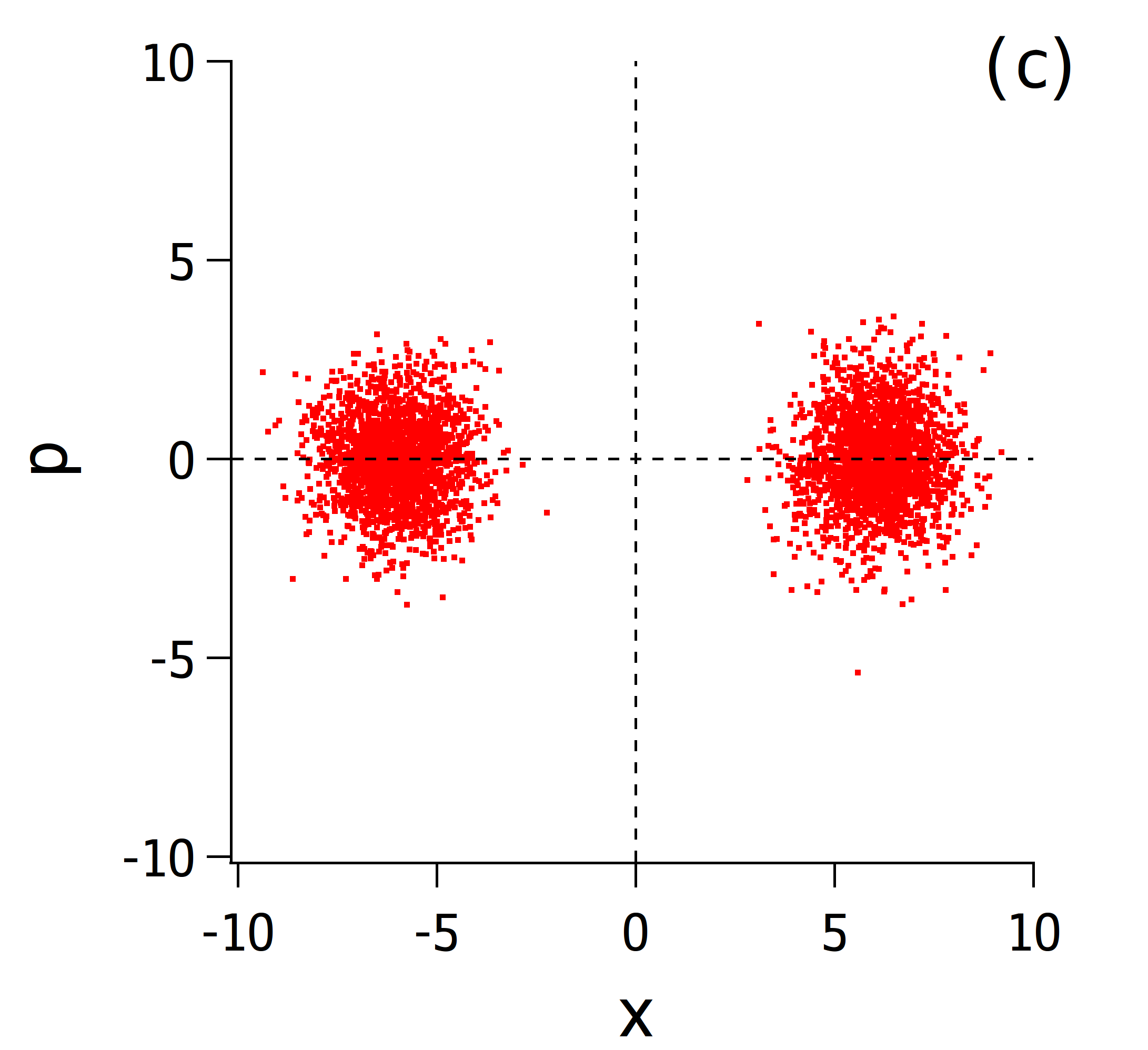}
\caption{(a) Measurement data for 50 photon signal with no phase correction. No bit clusters are apparent. 50 photon signal level is selected to make the clusters distinguishable for illustration in following sub figures. (b) Phase corrected using reference pulse measurements revealing distinct, but ambiguous, bit clusters. (c) Final phase correction using pattern matching reveals bit 0 and bit 1 clusters.}
\label{figPhaseCorrection}
\end{figure}

Alice generates two light pulses from a common coherent source and sends each to Bob, see Fig. \ref{alicePulses}. The first is the \emph{signal} pulse, having $\approx$ 1 photon and carrying the phase encoding. The second is the \emph{reference} pulse, having a larger photon number, $\approx$ 1000, and is responsible for aligning the relative phase between Alice and Bob's LO. In general, Bob receives Alice's pulses and performs a conjugated homodyne measurement with respect to his LO. After scaling for the overall transmittance, this provides the quadrature values ($X_{B},P_{B}$), which are related to Alice's quadratures values ($X_{A},P_{A}$), by the transformations

\begin{align}
X_B\left(\theta\right) &= X_A \cos\theta + P_A \sin\theta + N_X \\
P_B\left(\theta\right)&= - X_A \sin\theta + P_A \cos\theta + N_P
\end{align}
where $\theta$ is the relative phase between Alice's specific pulse and Bob's LO. $N_X$ and $N_P$ are the zero mean Gaussian noise for each quadrature. Measurements of $X_B$ and $P_B$ on Alice's signal with no phase corrections results in no apparent encoding as seen in Fig. \ref{figPhaseCorrection} (a). Phase correction is necessary to extract Alice's original encoding and is performed using the reference pulse.

Let us define Alice's reference pulse as having phase $\theta=\phi$ relative to Bob's LO. First we must determine this rapidly changing phase $\phi$ between Bob's LO and Alice's reference pulse. The reference pulse has no encoding, so we define the original quadrature position of Alice's reference pulse to have $P_A=0$ then 

\begin{align}
X_R=X_B\left(\phi\right) &= X_A \cos\phi + N_X\\
P_R=P_B\left(\phi\right)&= - X_A \sin\phi + N_P\label{XP_2}
\end{align}

where we assume $X_A >> N_X$ and $X_A >> N_P$ allowing approximate phase extraction by:

\begin{equation}
\phi = -\arctan{\frac{P_R}{X_R}}\textrm{.}
\label{phi}
\end{equation}

Alice's signal pulse has phase $\theta=\phi+\delta$, where $\delta$ is a constant phase offset between Alice's reference and signal pulses \footnote{The phase $\delta$ is actually slowly-varying in time, but it is approximately constant over the time-scale of a single-packet, $<$10 ms, which is all that is required.}. Once $\phi$ is determined from Eq. \ref{phi} we can determine the signal quadratures using the first phase correction, then by transforming into Bob's rotated quadrature basis ($X_{S}',P_{S}'$):

\begin{align}
X_S&=X_B\left(\phi+\delta\right)\nonumber\\
&= X_A \cos\left(\phi+\delta\right) + P_A \sin\left(\phi+\delta\right) + N_X \\
P_S&=P_B\left(\phi+\delta\right)\nonumber\\
&=- X_A \sin\left(\phi+\delta\right) + P_A \cos\left(\phi+\delta\right) + N_P
\end{align}

\begin{align}
X_S' &= X_S \cos\phi - P_S \sin\phi \\
P_S' &= X_S \sin\phi + P_S \cos\phi \textrm{.}
\end{align}

which results in

\begin{align}
X_S' &= X_A \cos\delta + P_A \sin\delta + N_X\\
P_S' &= -X_A \sin\delta + P_A \cos\delta + N_P\textrm{.}
\label{almostCorrected}
\end{align}

The phase $\delta$ is a constant over a single CV-QKD packet ($<$10~ms) which contains 8k distinct signal and reference pulses. Our GMCS protocol packets include headers that are 128 binary-phase encoded symbols.  After phase correction using the reference phase, the header's 0 and 1 encoding measurements are as distinguishable as they would be with $\delta = 0$, but it is ambiguous which cluster represents 0 or 1, see Fig. \ref{figPhaseCorrection}(b). Once the $\delta$ is known the phase-recovery is complete with unambiguous 0 and 1 encoding clusters as seen in \ref{figPhaseCorrection}(c). These same phase corrections are applied to the entire 8k symbol packet, the GMCS encodings, completing the phase correction process. We address determination of $\delta$ later in Section \ref{packetStructure}.

\section{Experiment}

\subsection{Light Generation, Modulation, and Detection}\label{experiment}
Alice's CV-QKD transmitter and Bob's receiver are depicted in Fig. \ref{AliceExpt} and \ref{BobExpt}, respectively. Both Alice and Bob utilize a frequency-stabilized continuous wave laser (OE Waves OE4028) to generate 1542 nm light for CV-QKD packet transmission (Alice) and LO generation (Bob) within their respective apparatus. Alice utilizes a LiNbO$_{3}$ waveguide intensity modulator (EOSpace) to generate 12 ns-wide pulses at a 1 MHz repetition rate. These pulses pass through polarization-maintaining optical fiber \footnote{Alice's transmitter and Bob's receiver utilize all polarization maintaining fiber. The link between them is standard single-mode fiber that does not maintain the polarization.}, beam-splitters, attenuators, and combiners to generate reference and signal pulses with orthogonal polarization containing approximately 1000 photons and $<$10 photons, respectively. Alice's configuration implements the phase noise reducing design proposed in \cite{marie2017self}. 
\begin{figure}[tb]
    \centering
    \includegraphics[width=0.6\linewidth]{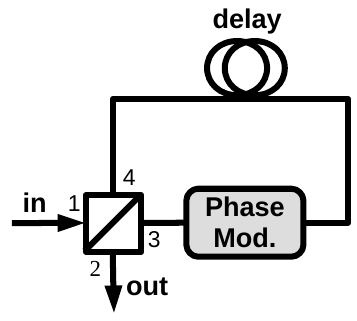}
    \caption{A LiNbO$_{3}$ phase modulator asymmetrically placed within a Sagnac fiber loop, allows Alice to encode phase and amplitude ($X,Y$) with a single device. The fiber loop is 20 m, approximately 100 ns of delay.}
    \label{amplitudePhase}
\end{figure}

Within Alice's signal path a bias-free phase-amplitude modulator \cite{dennis1996inherently,qi2018noise,Zhao:22} is utilized to apply the required pulse attenuation and phase needed to encode her quadratures $X,P$. For the GMCS encodings, each encoding is a random Gaussian distributed value based on random bits from a quantum random number generator (QRNG - ID Quantique).  In our experiment, the primary benefit of this modulator design is that it requires only a single digital-to-analog converter (DAC) compared to the two required when using two separate modulators. As seen in Fig. \ref{amplitudePhase}, asymmetrically placing a LiNbO$_{3}$ phase modulator (EOSpace) within a Sagnac loop allows different phases to be applied to the clockwise and counter-clockwise paths in the 20 m loop, a 100 ns delay. When light is only incident at the input, it is straightforward to show that the coherent state at the output is
 \begin{equation}\alpha_{\textrm{out}}=  e^{i\frac{\left(\phi_{\textrm{cw}} + \phi_{\textrm{ccw}}\right)}{2}}\sin\left(\frac{\phi_{\textrm{cw}}-\phi_{\textrm{ccw}}}{2}\right)\alpha_{in}\textrm{,}\end{equation}
 where $\phi_{ccw}$ and $\phi_{cw}$ are the phases applied to the counter-clockwise and clockwise propagating pulses, see Appendix \ref{phase-amplitude}. Thus, adjusting the phase difference $\phi_{ccw} - \phi_{cw}$ adjusts the light amplitude while the sum $\phi_{ccw} + \phi_{cw}$ adjusts the global phase. 
 
 A wavelength division multiplexer (WDM) is used to combine the CV-QKD packet generated by Alice's optical scheme above, with the classical communications channels from a 1270 / 1330 nm bi-directional optical transceiver (BiDi) prior to being transmitted to Bob.  Thus, the fiber carries two classical channels and a single quantum channel.
\begin{figure}[b]
\includegraphics[width=0.9\linewidth]{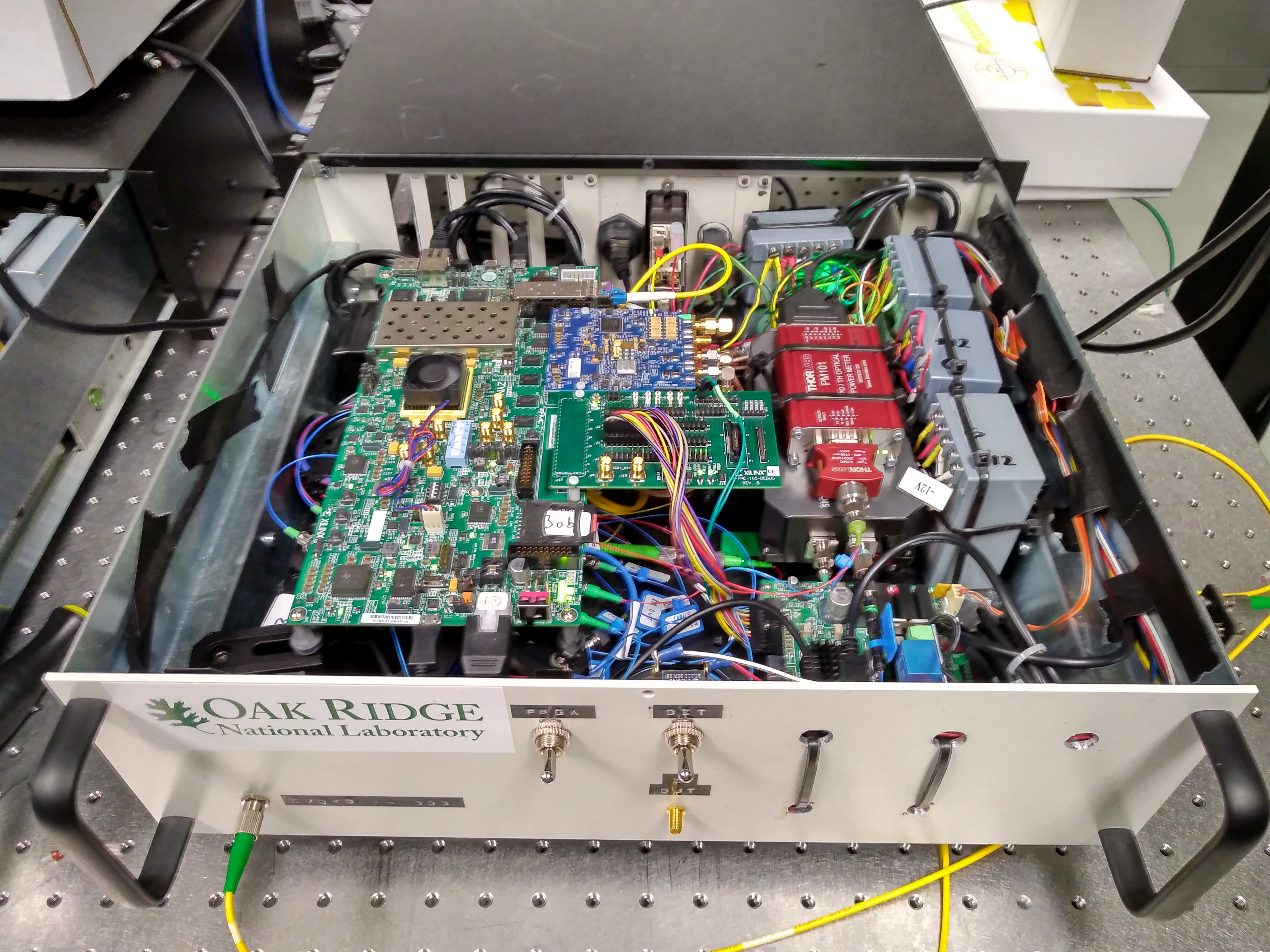}
\caption{Bob's CV-QKD receiver (pictured) and Alice's transmitter are both rack-mountable units which include all system components.}
\label{box}
\end{figure}

Using the same model of laser and modulator as Alice, Bob generates LO pulses 12 ns in duration and also uses fiber splitters, attenuators, and combiners to realize a LO for the signal and a LO for the reference with orthogonal polarizations, each with 8 $\mu$W of average power. Bob's repetition rate is also 1 MHz (like Alice), however, he generates 4 total pulses per 1 $\mu$s period counting the additional clone pulses, see Section \ref{biasCorrection}. A WDM filter at Bob's input splits the CV-QKD packets from the classical communications channels. The CV-QKD packets undergo polarization correction to align the signal and reference pulses prior to detection. Bob employs an Optoplex $90^\circ$ optical hybrid balanced detector to perform heterodyne detection. Bias controllers (PlugTech MBC-NULL-03) are used on both Alice and Bob's intensity modulators to maintain a null output - providing an extinction ratio of 30 dB when the full modulator range is used.

\begin{figure*}
\includegraphics[width=0.7\linewidth]{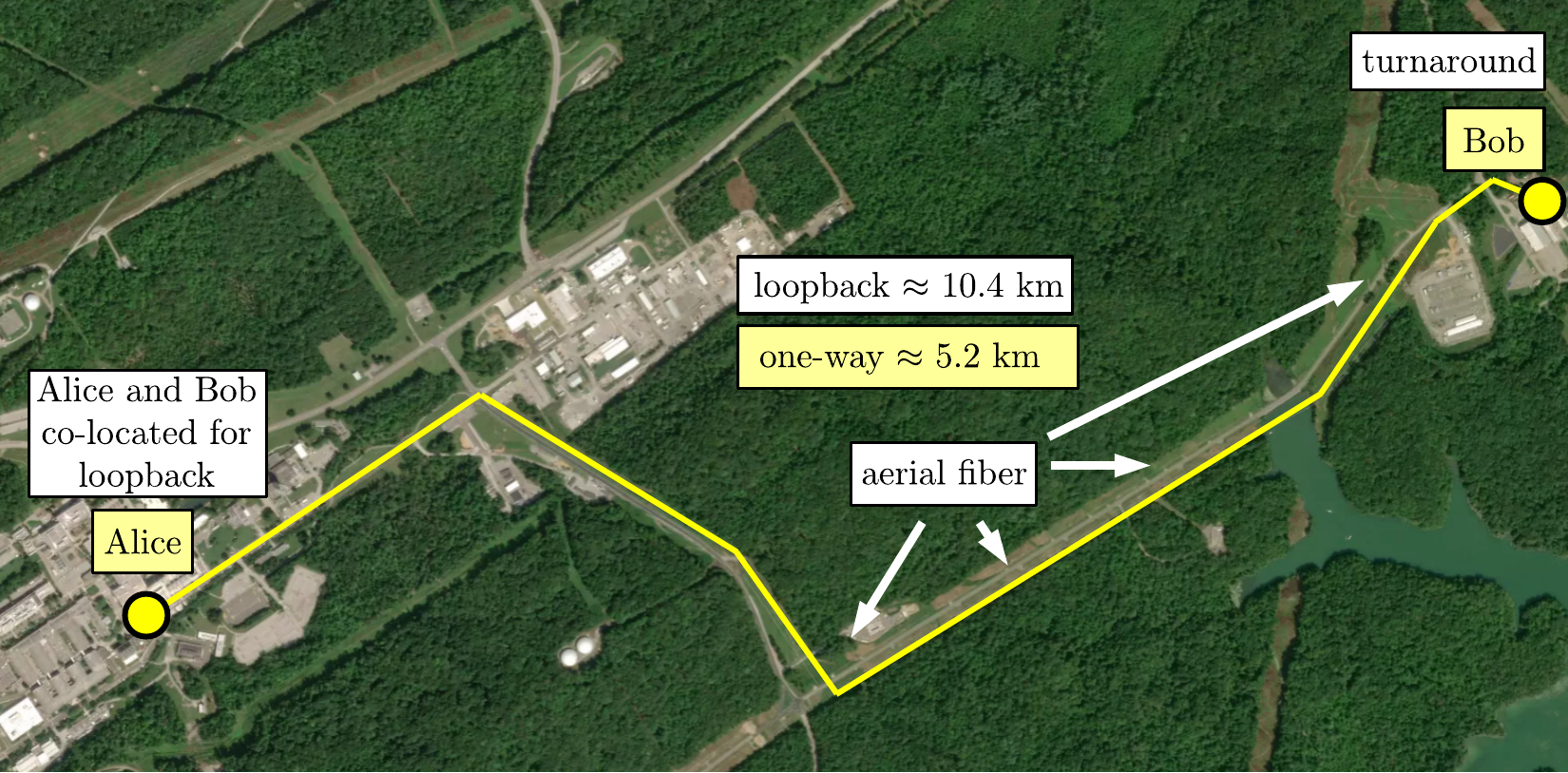}
\caption{Aerial photo of Oak Ridge National Laboratory depicting approximate optical fiber path from Alice to Bob in which quantum and classical communications coexist. We demonstrated operation of the CV-QKD system over a 10.4 km loop-back link with Alice and Bob co-located in the same laboratory and also a point-to-point link in which Alice and Bob are separated spatially by 3 km and linked by 5.2 km of optical fiber.}
\label{figPath}
\end{figure*}

Alice's transmitter and Bob's receiver both contain field-programmable gate arrays (FPGAs) (Xilinx Zynq-7000 ZC706 with Analog Devices FMCDAQ2 daughter cards) which provide two high-speed digital to analog conversion (DAC) outputs and two analog to digital conversion (ADC) inputs per node. Alice and Bob both use the DAC outputs with inline amplifiers (Mini-Circuits ZX60-100VH+) to drive their respective modulators. Bob utilizes the ADC inputs to record heterodyne detector measurements. Bob also uses low-speed digital outputs via a daughter card (FMC XM105) to drive the polarization controller (General Photonics PolaRite III). All the components of Alice and Bob's systems fit in rack-sized enclosures as seen in Fig. \ref{box}.

\subsection{Coexistent Quantum and Classical Communications}
Classical communications between Alice and Bob are performed within the same optical fiber strand on which Alice's quantum signals are transmitted. This is accomplished using bi-direction (BiDi) small form-factor pluggable (SFP) optical transceivers operating at 1270 nm and 1330 nm for Alice-to-Bob and Bob-to-Alice transmissions, respectively. The classical communications are multiplexed with the quantum signal using a 1310 nm / 1550 nm fiber-based splitter combiner with 0.2 dB insertion loss and approximately 25 dB of isolation for each channel. The optical transceivers used are designed for 10 km distance range and each transmits approximately 600 $\mu$W of optical power in the O-band. An additional 5 dB attenuator is used at Alice's BiDi to limit the optical power to the minimum required to complete the optical link. Bob uses two 1310 nm / 1550 nm fiber splitters in series to further improve the isolation between the classical and quantum channels before signal detection.

\subsection{The GMCS Protocol and Performance Metrics}

The performance of the GMCS CV-QKD system is evaluated, in part, based on it's secret key rate which is determined from CV-QKD system parameters, specifically the line transmission $T$ between Alice and Bob, the excess noise $\xi$ in Alice's signal before transmission, the detector efficiency $\eta$, and the detector noise $\upsilon_e$, see Appendix \ref{secretKey}. These parameters are specifically relevant because they describe the signal created by Alice, detected by Bob, and potentially intercepted by a notional adversary Eve.

Independent of the theoretical constructions necessary to generate a security proof, the practical implementation of the GMCS protocol proceeds as follows. Alice derives each of her encoded quadratures from independent random Gaussian distributions with variance $V_A$. She then sends the encoded states to Bob via a channel with transmission $T$. The state encoding comes with an excess noise $\xi$ which for security must be attributed to Eve accessing the encoded states during transmission. Bob utilizes heterodyne detection to measure the transmissions with a fixed loss of 0.5, detector efficiency of $\eta$, and detector noise $\upsilon_e$.

The total variance in shot-noise units for a single quadrature of Alice's encoding and excess noise is
\begin{equation}V_{A-\textrm{total}}=V_A + \xi\textrm{.}\end{equation}
Bob measures
\begin{equation}V_{B}=\frac{T\eta}{2}\left(V_A + \xi\right) + 1 + \upsilon_e\label{BobMeasures}\end{equation}
where the additional factor of 1 is from the shot-noise present in one quadrature of Bob's heterodyne detection.  

\subsection{CV-QKD Demonstration}
\begin{figure}[tbh]
\includegraphics[width=0.8\linewidth]{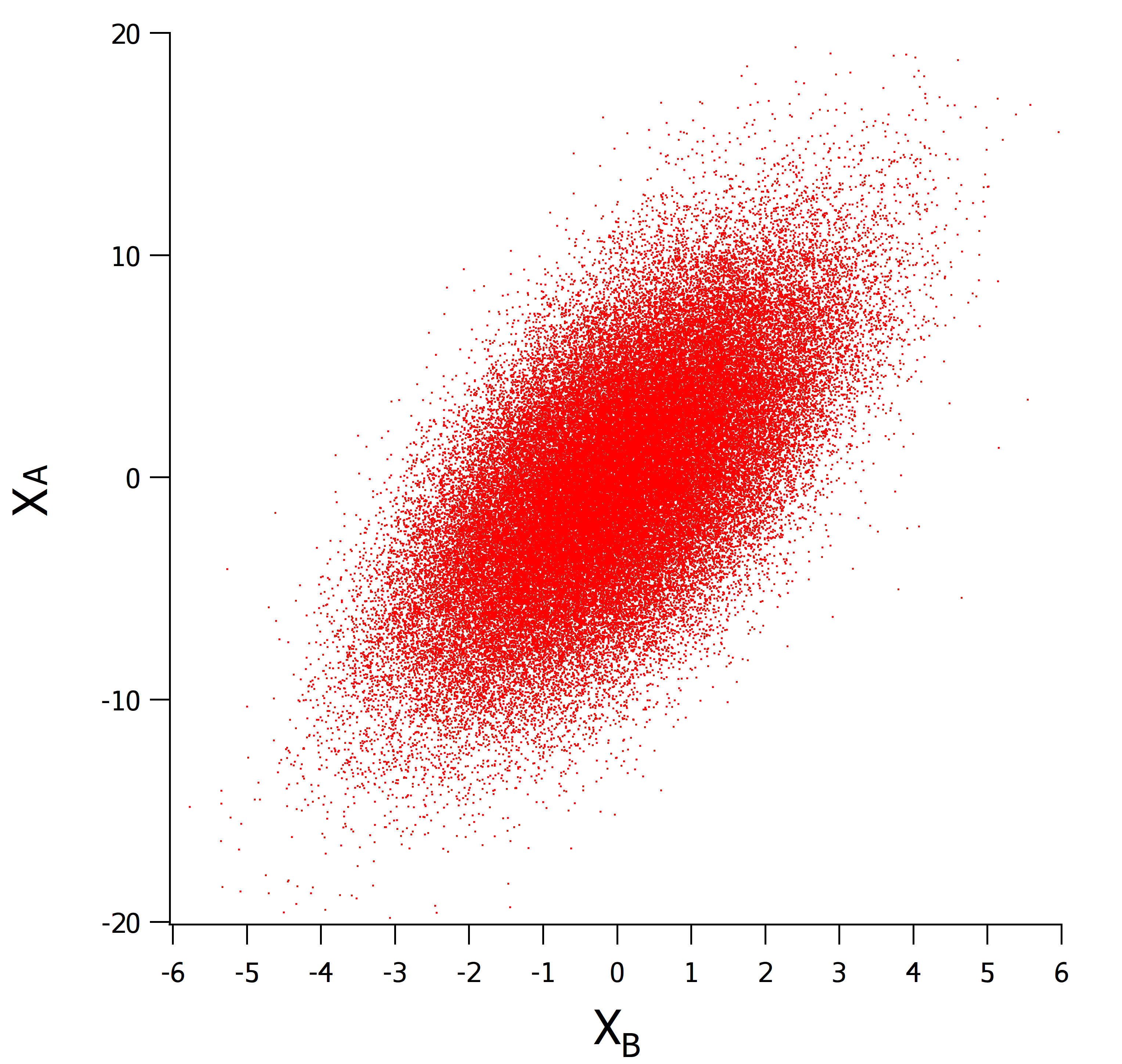}
\caption{Alice's encoded $X_A$ values from a Gaussian distribution with $V_A = 25$ versus Bob's measured $X_B$ values, in SNU. The smaller $X_B$ values result from transmission loss, detection inefficiency, and heterodyne detection, 15 dB total loss.}
\label{correlation}
\end{figure}

We performed a CV-QKD demonstration over an approximate 10.4 km loop-back link, which has 8 dB of loss, shown in Fig. \ref{figPath}. The two fiber links making up the loop-back link are distinct fibers. In the loop-back test, Alice and Bob were co-located in the same laboratory though there was only a single 10.4 km optical fiber link between them which they utilized for both quantum and classical communications--there was no electronic link. With multiplexed quantum and classical signals, Alice and Bob generated raw key using a GMCS protocol with CV-QKD packets containing 8k signal pulses. The symbol rate for the system is 50 kSymbols/s, 5$\%$ of the rep-rate, the reduction due to processing and packet-handling overhead. The symbol rate was determined by sending 100 packets, 8k signals per packet, in succession and timing the total time between the start of the first packet to the end of the final packet's post-processing. An example of the correlations between Alice's GMCS $X_A$ encodings and Bob's measured $X_B$ values are shown in Fig. \ref{correlation}.  

\begin{figure}[t]
\includegraphics[width=\linewidth]{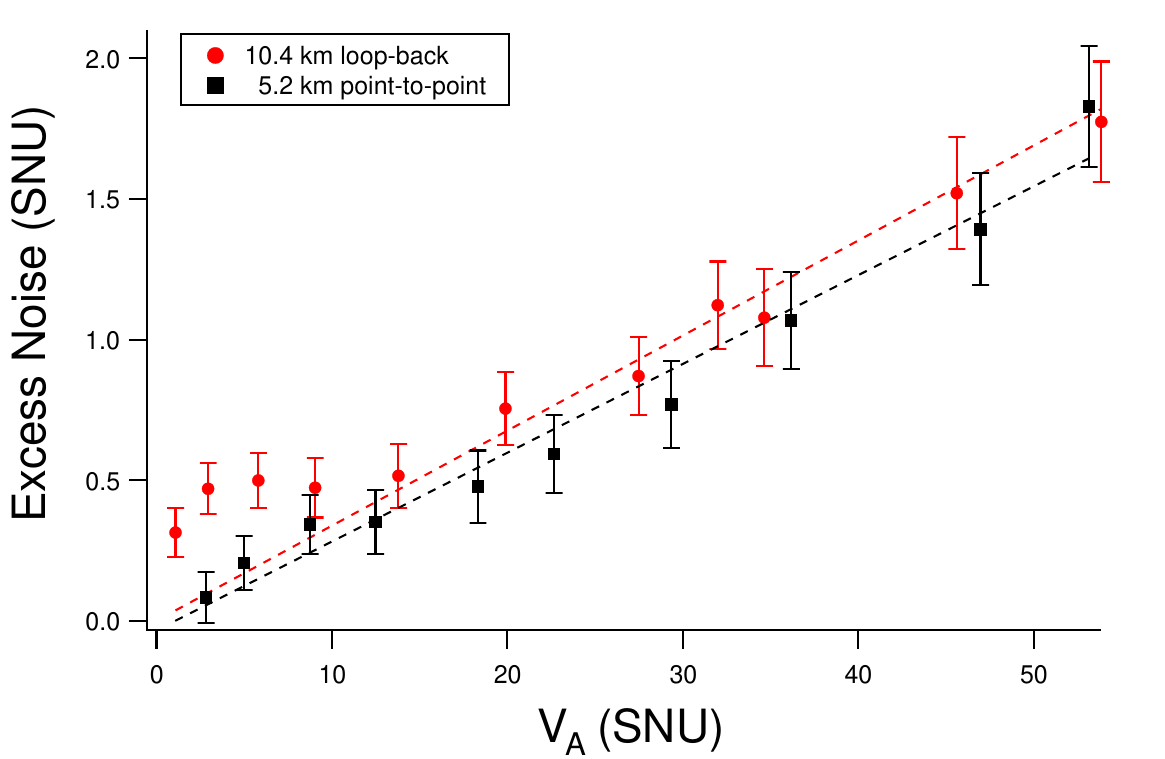}
\caption{Excess noise versus Alice's modulation variance $V_A$ for 10.4 km loop-back and 5.2 km point-to-point tests. We determine excess noise values for 8k symbol packets which we average over 100 packets for a final estimate at each $V_A$. Line fits assume a phase-noise model.}
\label{excessNoise}
\end{figure}

To further validate our system, we performed a point-to-point test in which Alice and Bob are physically separated by 3 km and linked by a 5.2 km deployed optical fiber with 3.2 dB of loss, see Fig. \ref{figPath}. In this test, we logged into Alice's Zynq SoC Linux operating system (OS) to execute the CV-QKD transmitter software. From Alice's OS we remotely logged into Bob's OS and executed the CV-QKD receiver software utilizing the classical communications link coexistent with the quantum signals. As seen in Fig. \ref{excessNoise}, the excess noise versus Alice's modulation variance $V_A$ is similar in the loop-back and point-to-point tests. We note that the shorter point-to-point test data is more consistent, we speculate that this is due to the lower loss link (stronger signal).   

Alice's true encoding value $X_A$ in shot noise units (SNU) is related to her digital encoding value $\mathds{X}_A$ by a constant
\begin{equation}X_A = k\mathds{X}_A\textrm{,}\end{equation}
with a similar relation for her $P$ encodings. Given Alice's digital encoding values and Bob's measured data we can use linear regression analysis to determine $k$ since
\begin{equation}
X_B = k \sqrt{\frac{\eta T}{2}}\mathds{X}_A + X_0
\end{equation}
where $X_0$ is a Gaussian random variable (noise) which averages to zero. Given the known transmission $T$ and detection efficiency $\eta$, it is straightforward to determine $k$, $V_A$, and $\xi$. Since we know exactly the variance $\mathds{V}_A$ of the digital encoding $\mathds{X}_A$, then determining $k$ through linear regression gives us $V_A$
\begin{equation}V_A = k^2\; \mathds{V}_A\textrm{.}
\end{equation}
Finally, with $V_A$ known, we use Eq. \ref{BobMeasures} to determine $\xi$.
In a fully mature system, not a field-test, Alice would know her true $V_A$ exactly and Bob would use this to estimate the transmission parameter $T$.

System noise - namely electrical and phase noise - has the impact of reducing Alice and Bob's secret key rate. Electrical noise includes all the noise present when no light (no signal and no LO) is introduced into the 90 degree optical hybrid detector, and is measured to be approximately $\nu_e = 0.175$ in shot-noise units. The optical phase recovery process is imperfect due to noise in the reference phase measurement and imperfections in the encoding and measurement processes. Phase uncertainty leads to an additional noise term in GMCS \cite{marie2017self}, given by
\begin{equation}\xi_\textrm{phase}=V_A \Delta\phi\label{phasenoise}\end{equation}
where $V_A$ is Alice's modulation variance and $\Delta\phi$ is the phase variance.  If we assume the excess noise in our system, shown in Fig. \ref{excessNoise}, is solely due to this phase noise, for our 10.4 km loop-back and 5.2 km point-to-point experiments we estimate $\Delta\phi = 0.034 \pm 0.002$ and $\Delta\phi = 0.030 \pm 0.002$, respectively. We note that the excess noise parameter $\xi$ is estimated without regard for it's origin. However, identifying the source of excess noise is the first step in reducing it.
\begin{figure}[t]
\includegraphics[width=\linewidth]{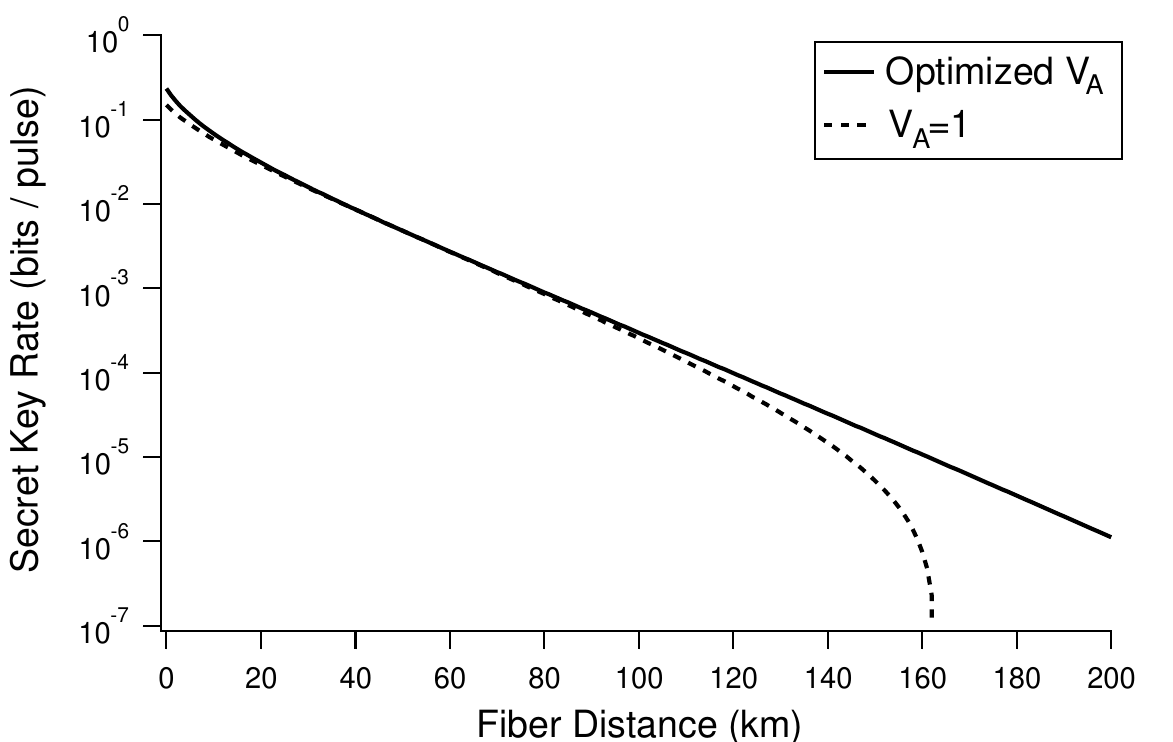}
\caption{ Secret key rate for a trusted receiver given measured and estimated system parameters.}
\label{SKRrange}
\end{figure}
\begin{figure}
\includegraphics[width=\linewidth]{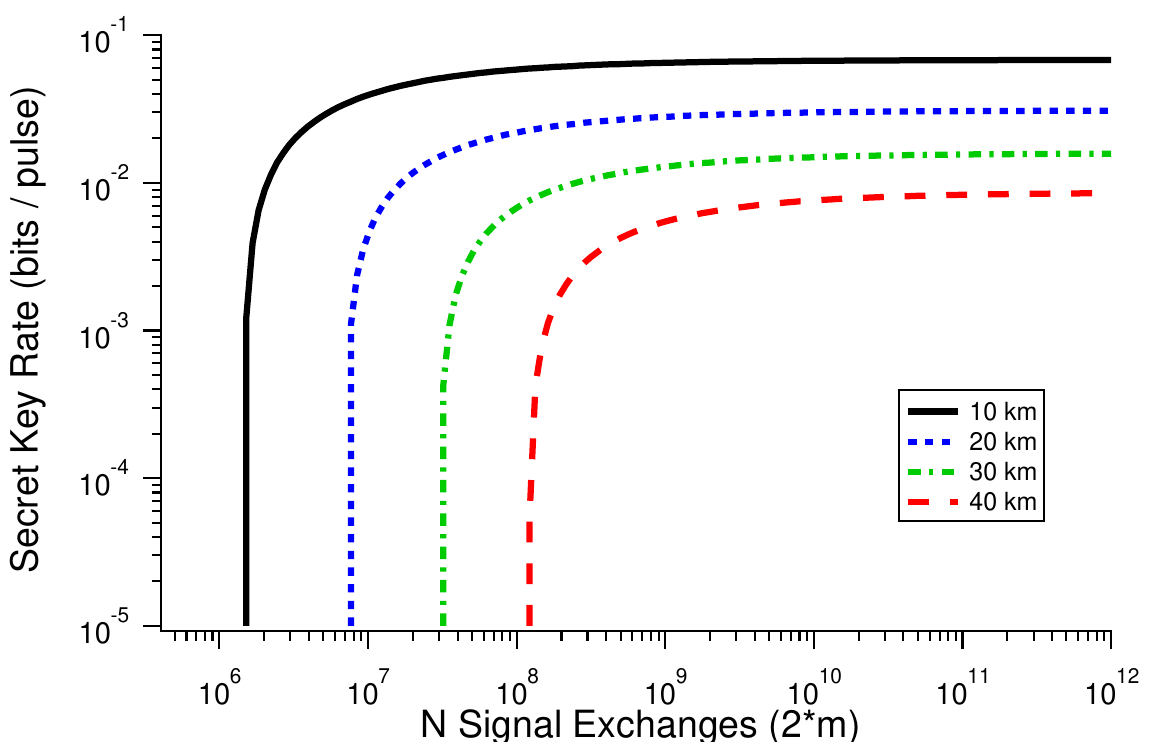}
\caption{ Secret key rate for a trusted receiver considering finite-size effects for 10 km, 20 km, 30 km, and 40 km transmission scenarios. Here was assume half the signal exchanges are used for parameter estimation.}
\label{SKRfinite}
\end{figure}

The theoretical expected secure key rate using a GMCS protocol is given by
\begin{equation}
R_{\infty}= \beta I_{AB}-\chi_{EB}
\label{R2}\end{equation}
and can be determined by calculating Alice and Bob's mutual information $I_{AB}$ and the Holevo information $\chi_{EB}$, a bound on Eve's potential knowledge. These quantities are determined using methods given in Appendix \ref{secretKey} which reproduces previous work \cite{lodewyck2007quantum,fossier2009improvement}. The primary factors contributing to the rate are the excess noise $\xi$, Alice's modulation variance $V_A$, the electrical noise $\nu_e$, and the channel transmission $T$. Our estimated secret key-rate versus distance is given in Fig. \ref{SKRrange} using an electric noise $\nu_e=0.175$, measured phase noise $\Delta \phi=0.034$, reconciliation efficiency $\beta = 0.95$, and effective detector efficiency $\eta=\eta_{\textrm{det}}T_{\textrm{Bob}}=0.42$ where the true detector efficiency is  $\eta_{det}=0.5$ and the transmission through Bob's receiver is $T_{\textrm{Bob}}=0.84$. The secret key-rates are given for both optimized modulation variance $V_A$ and constant $V_A=1$.

Parameter estimation is used to determine values such as the transmission and channel noise. The amount of data used in these estimations affects their uncertainty. Security dictates a conservative interpretation of this uncertainty. To take these finite-size effects into account we utilize estimation adjustments as detailed in Appendix \ref{secretKeyFS} which utilizes the methods in \cite{leverrier2010finite}. In brief, the first consequence of these finite-size considerations is a halving of the key-rate since we must utilize half the data to make parameter estimations. Second, there is a key-rate dependence on the total signal exchanges used to estimate parameters. We summarize our results in Fig. \ref{SKRfinite} which shows that secret key-rates for 10, 20, 30, and 40 km asymptotically approach a maximum value as more signal exchanges are utilized. We have used the same key parameter values as above for $\nu_e$, $\xi$, $\beta$, and $\eta$.

Experimental results from our CV-QKD system indicate that, accounting for finite size effects, the secure key rates given in Table \ref{table} are achievable assuming a 50 kSymbol/s data rate. We assume half the signals exchanged are used for parameter estimation. Increasing the repetition rate is one-way to improve the secure key rate.
\begin{table}[t]
\begin{center}\begin{tabular}{|c|c|c|} 
 \hline
 Distance & Key Rate & Data Collection\\ [0.5ex] 
 (km) & (kbps) & Time (hrs)\\ [0.5ex] 
 \hline
 10 & 1.6 & 5\\ 
 \hline
 20 & 0.74 & 20.8\\
 \hline
 30 & 0.38 & 80\\
 \hline
 40 & 0.20 & 302\\
 \hline\end{tabular}
\end{center}
\caption{Distributed final-key rate prediction for our system as a function of distance considering finite-size effects assuming a 50 kSymbol/s transmission rate.}
\label{table}
\end{table}
\section{Solutions to True LO Implementation Challenges}
There are several challenges to experimental implementation of the true LO protocol, in particular detector bias correction, timing synchronization, polarization correction, vacuum noise measurement, and signal-reference relative phase correction. We detail solutions to each challenge below.

\subsection{Detector Bias Correction\label{biasCorrection}}
\begin{figure}[b]
     \centering
     \begin{subfigure}[b]{0.49\linewidth}
         \centering
         \includegraphics[width=\linewidth]{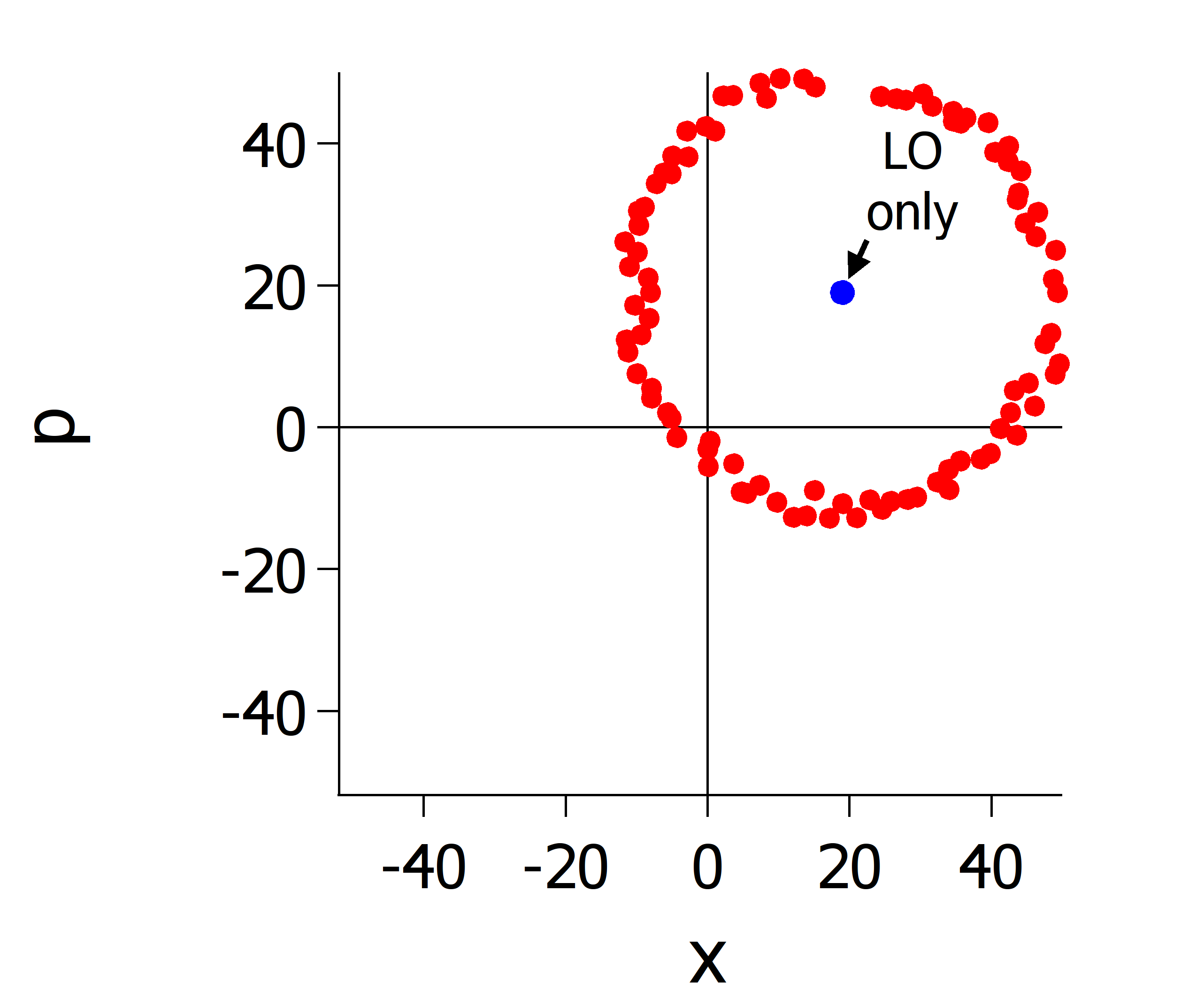}
         \caption{}
         \label{biasA}
     \end{subfigure}
     \hfill
     \begin{subfigure}[b]{0.49\linewidth}
         \centering
         \includegraphics[width=\textwidth]{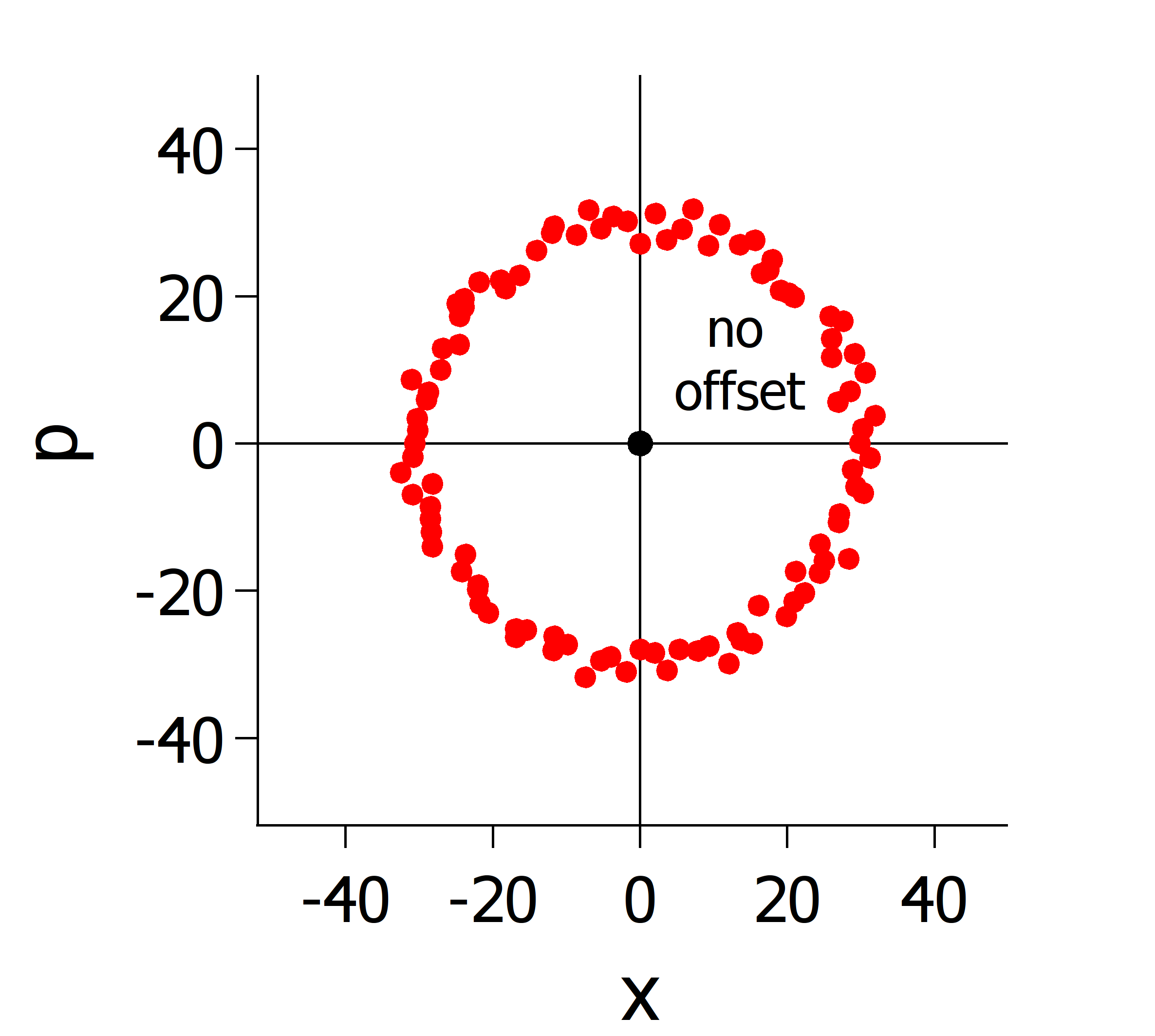}
         \caption{}
         \label{biasB}
     \end{subfigure}
     \newline
     \begin{subfigure}[b]{0.49\linewidth}
         \centering
         \includegraphics[width=\linewidth]{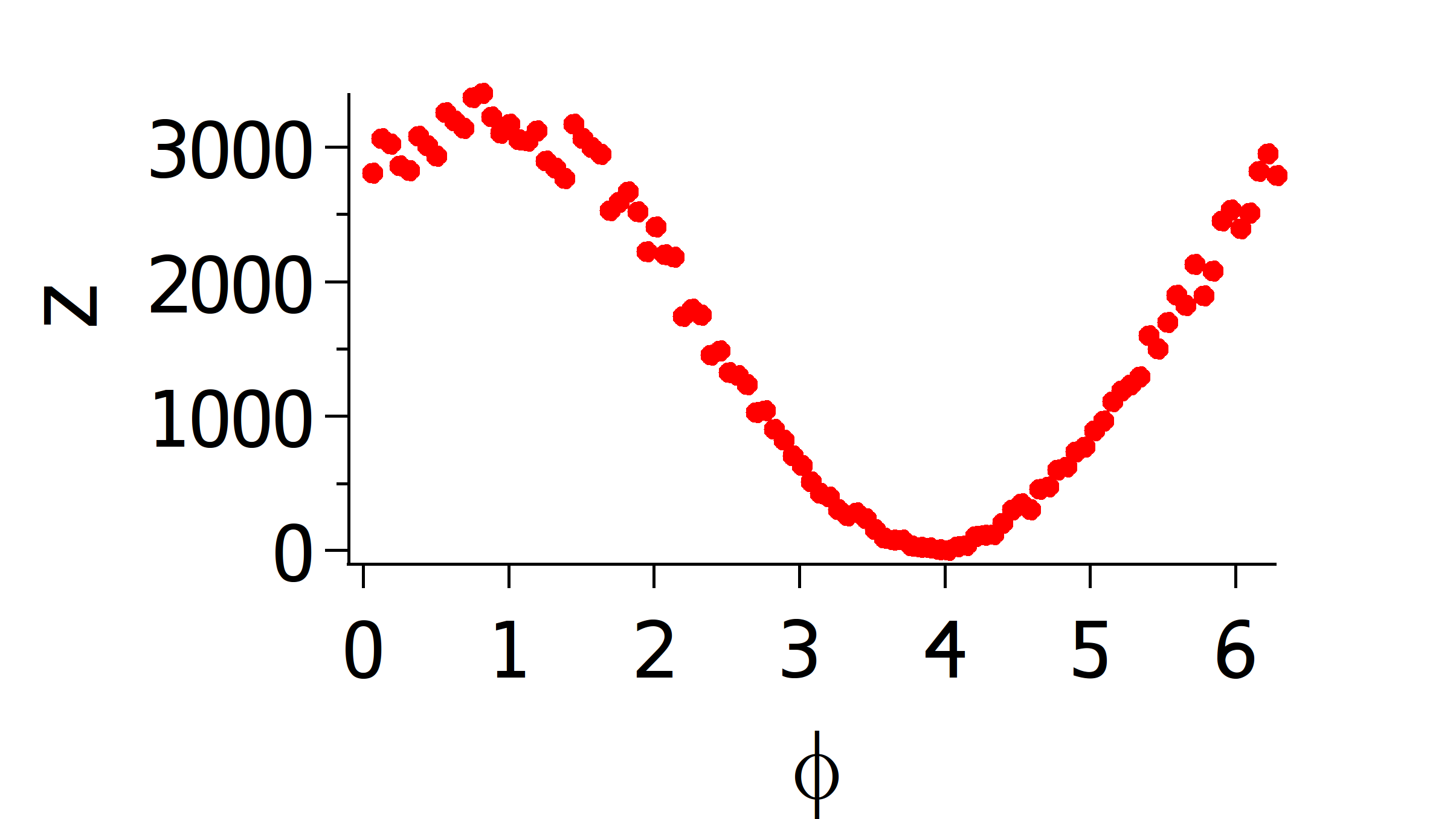}
         \caption{}
         \label{biasC}
     \end{subfigure}
     \hfill
     \begin{subfigure}[b]{0.49\linewidth}
         \centering
         \includegraphics[width=\linewidth]{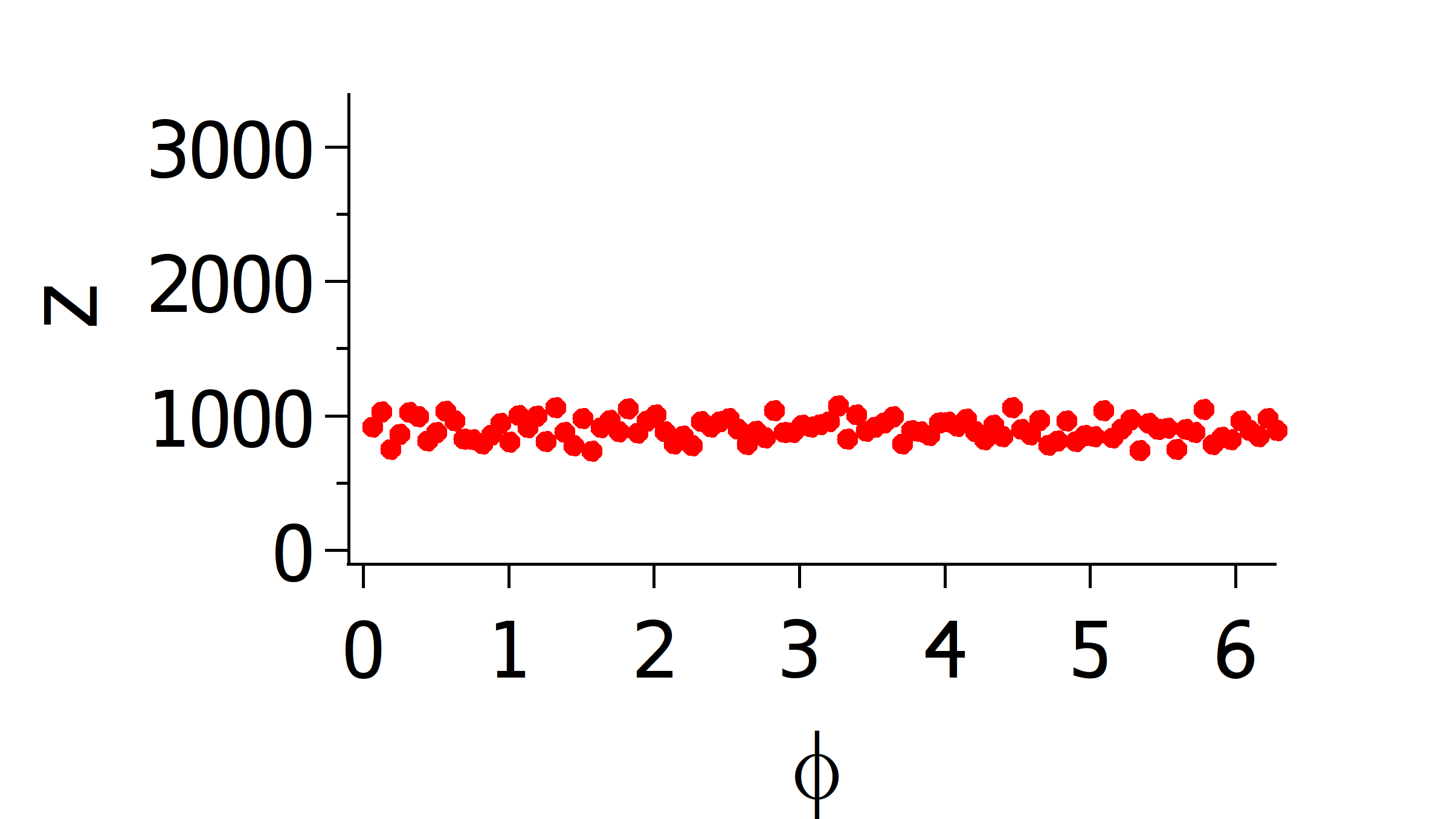}
         \caption{}
         \label{biasD}
     \end{subfigure}        
     \caption{Comparison of quadrature values $X,P$ measured at some instant in time and the resultant $Z=X^2+P^2$ values with and without bias correction. (a) Raw measured $X,P$ with bias offset. (b) Bias corrected $X,P$. (c) Raw measured $Z$ with bias offset. (d) Bias corrected $Z$.}
        \label{Bias correction}
\end{figure}

\begin{figure}[t]
     \centering
     ~\quad\begin{subfigure}[b]{\linewidth}
         \centering
         \includegraphics[width=\linewidth]{pulses.png}
         \label{biasA}
     \end{subfigure}
     \newline
    \begin{subfigure}[b]{\linewidth}
         \centering
         \includegraphics[width=\linewidth]{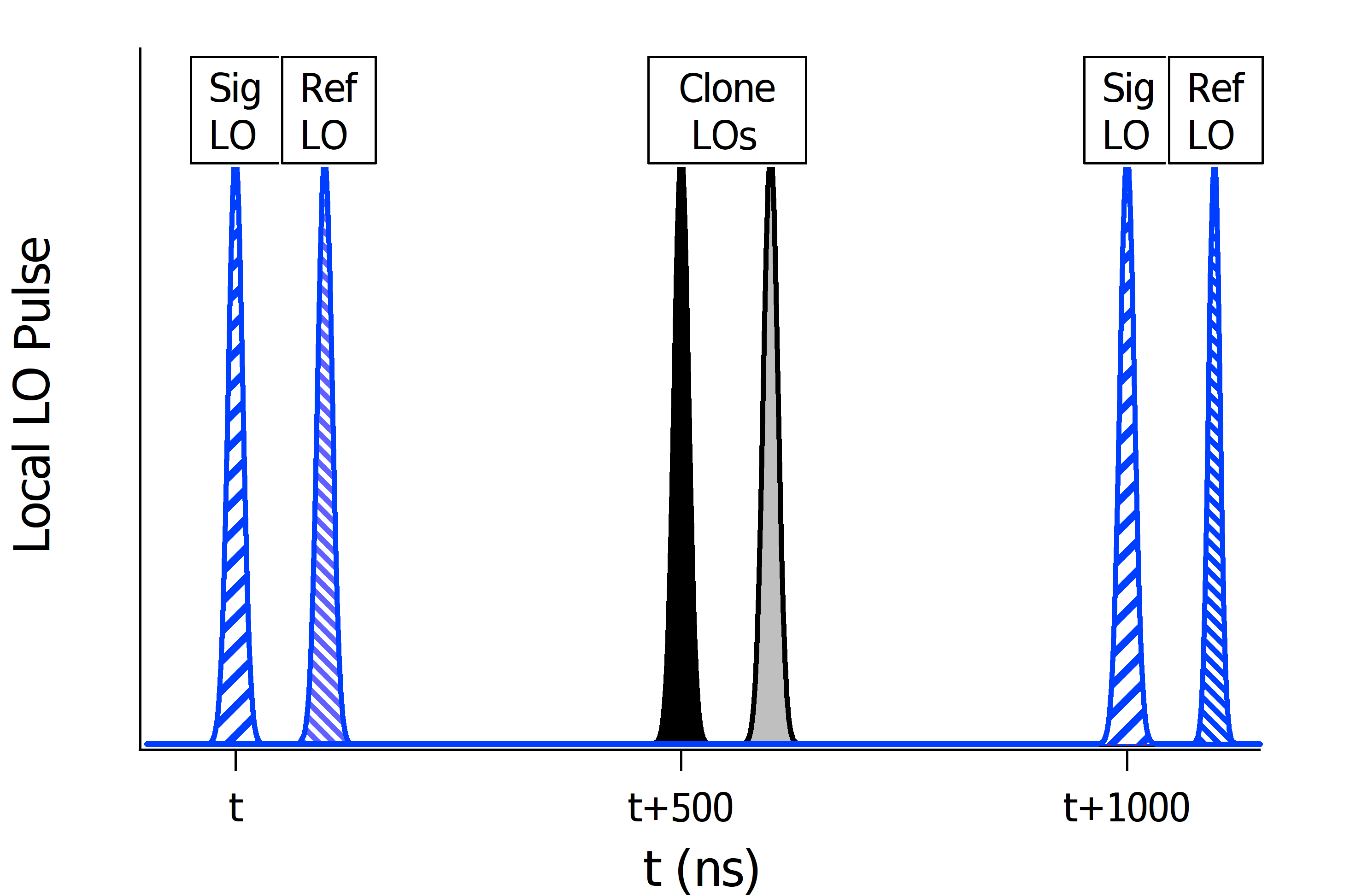}
         \label{biasB}
     \end{subfigure}
     \caption{(a) Alice reference and signal pulses versus time, not to scale. (b) Bob's LO and clone LO pulses versus time, not to scale.}
        \label{cloneLO}
\end{figure}

Ideally our CV-QKD measurement results would mirror the theory presented in Section \ref{Local LO CV-QKD}. Unfortunately, real devices have deficiencies that prevent perfect agreement. In our prototype device the 90-degree optical hybrid detector has an imperfect common-mode rejection ratio. In practical terms, this means that the origin of phase space data is not at zero, as illustrated in Fig. \ref{Bias correction} (a). The location of this origin depends on the specific detector bias and the power of the LO. The measured $X$ and $P$ values can be described as
\begin{align}
X_{meas}(t) &= X(t)+\alpha(t)\\
P_{meas}(t) &= P(t)+\beta(t)
\end{align}
where $\alpha(t)$ and $\beta(t)$ are the bias offsets in the $X$ and $P$ quadratures, respectively, which depend on the LO power. These offsets pose a problem in our system, as the detection of an origin centered on $X$ and $P$ is needed to determine an accurate $Z$ value. The $Z$ value is useful, since it is proportional to the received photon number. Specifically, for a given $X$ and $P$
\begin{equation}Z=X^2+P^2\end{equation}
and the detected photon number with heterodyne detection is
\begin{equation}n_{\textrm{det}}=\frac{Z-Z_{\textrm{shot}}-Z_{\textrm{elec}}}{ Z_{\textrm{shot}}}\end{equation}
where $Z_{shot}$ is the $Z$ contribution from shot-noise and $Z_{elec}$ is the contribution from all electrical sources when no light is incident into the detector. The photon number arriving at the heterodyne detector is $n=2n_{\textrm{det}}/\eta$ 
where $\eta$ is the detector efficiency. Monitoring of the reference pulse $Z$ value forms the basis for the timing synchronization, signal pulse heralding, and polarization correction. Fig. \ref{Bias correction} (a) and (c) illustrate the issue, where biased $X$ and $P$ values lead to a varying $Z$ value. By correcting the bias, as seen in Fig. \ref{Bias correction} (b) and (d), we are able to recover the required measurement. Performing the bias correction is itself a challenging issue since the LO is pulsed, we must match the bias contribution's amplitude in time. We require these corrections to take place on the FPGA using basic logic gates, but complex normalizing operations are challenging to implement. Our solution is for Bob to generate clone LO's after the original LO pulses, as illustrated in Fig. \ref{cloneLO}. Given that our true LO pulses are generated at time $t$ we generate clone LO pulses at $t+500$ ns. Since the pulse repetition period is $1 \mu$s, there are also LO pulses at $t-500$ ns. In other words, we are generating a 2 MHz LO signal whereas our signal from Alice is 1 MHz. Thus, every other LO pulse is the ``clone". The short time between these pulses does not allow for significant variation in laser power. Measurements in the clone bins contain no light present from Alice, only the bias offset from the clone LO.
\begin{align}
X_{meas}(t-500) &=\alpha(t-500)\approx \alpha(t)\\
P_{meas}(t-500) &=\beta(t-500)\approx\beta(t)\textrm{.}
\end{align} 
Data incoming to the FPGA is corrected by taking the difference of bins separated by $500$ ns.
\begin{align}
X(t) &= X_{meas}(t)-X_{meas}(t-500)\\
P(t) &= P_{meas}(t)-P_{meas}(t-500)\\
Z(t) &= X^2(t)+P^2(t)
\end{align}
This bias correction process adds noise, since we are now incorporating two LOs hence doubling the shot-noise. However, this bias-corrected $Z$ value is only used in the FPGA triggering and polarization correction algorithms. The uncorrected $X$ and $P$ values are also collected and bias-corrected in post-processing, thus avoiding additional noise from the clone LO.

\subsection{Timing Synchronization}
The operation of our CV-QKD system requires that Alice's light pulses and Bob's LO pulses arrive at Bob's 90$^\circ$ optical hybrid detector at the same time. To see Alice's reference pulse, Bob's LO pulse must be present, but to trigger the correct LO pulse generation time Bob must see Alice's reference pulse. To solve this, Bob always generates his LO at a 1 MHz rate, even when he detects no reference pulse from Alice. While Alice and Bob's FPGA clocks are set to the same frequency setting, 250 MHz \footnote{1 Ghz operation is achieved at the ADC/DAC daughter card level.}, they have a finite drift between them due to the distinct internal oscillators used in the separate FPGAs. This frequency drift results in eventual overlap of Alice's reference pulse and Bob's LO. This occurs rapidly after system start-up, since the drift must shift the pulse timing by, at most, 1 $\mu$s. Once a single overlap of Alice's reference pulse and Bob's LO occurs, Bob locks onto this time-bin and each subsequent reference pulse refreshes the lock. Bob's LO generation is restricted such that he may only shift his clock by 1 ns for each pulse period, 1 $\mu$s, as he attempts to maintain his lock to Alice's reference pulses. This restriction improves the locking performance and is also important to the bias correction scheme previously described Section \ref{biasCorrection}. If Bob loses his lock (e.g., by missing a reference pulse), he continues generating his LO at 1 MHz until he sees another reference pulse from Alice.

\subsection{Real-Time Vacuum 
Measurement\label{real-time}}
\begin{figure}[b]
\includegraphics[width=\linewidth]{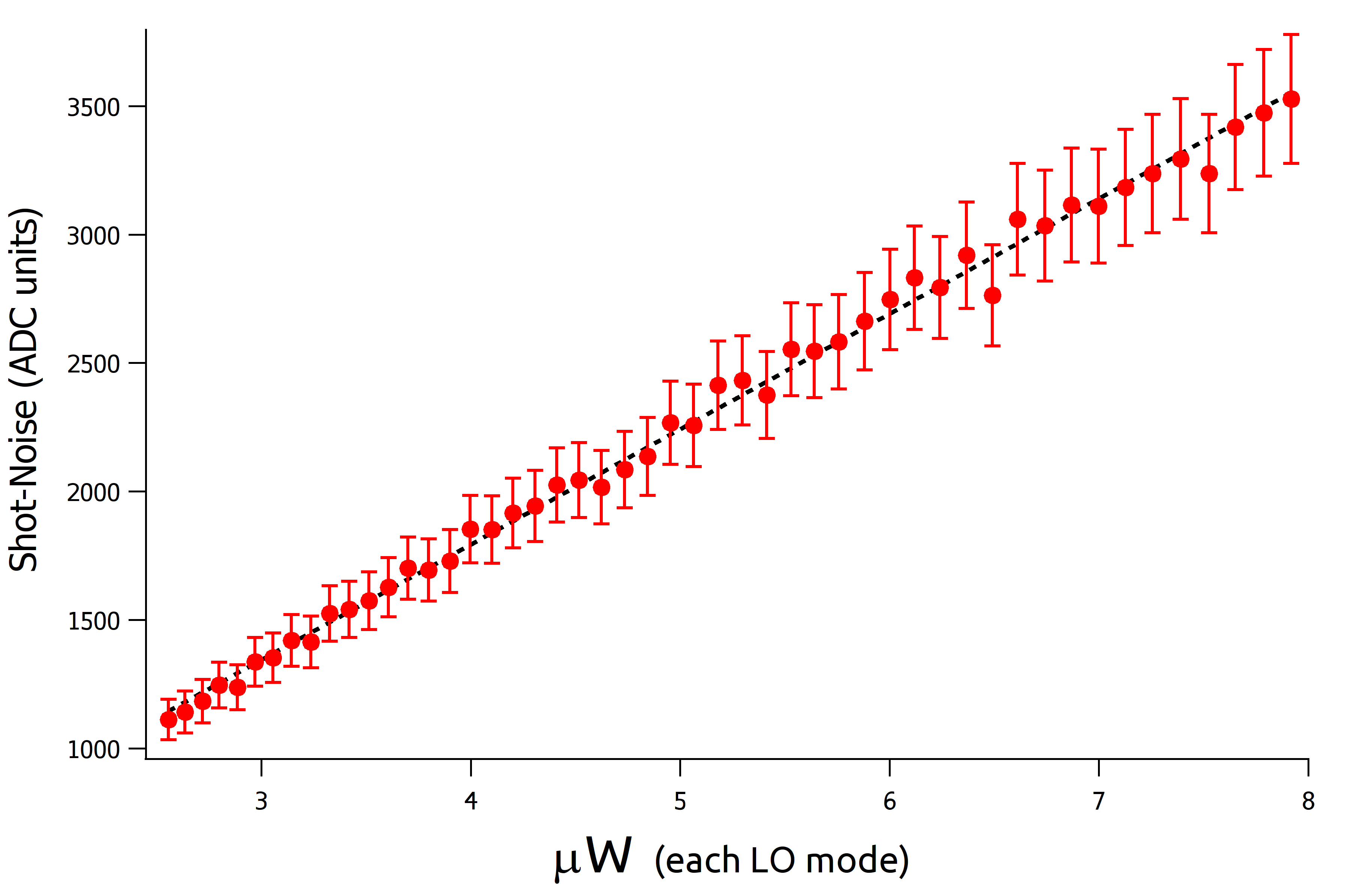}
\caption{Shot-noise measured versus the LO power. The shot-noise is measured in the signal clone LO time-bin when no light from Alice is present. As required, we see a linear relationship with the local oscillator (LO) power, the line-fit has $R^2=0.995$.}
\label{shot-noise}
\end{figure}
CV protocols typically rely on knowledge of the shot-noise which provides units of measure for all other experimental parameters. In CV-QKD, shot-noise measurement is critical since it impacts metrics for secret key rates. An incorrect measurement of the shot-noise could lead Alice and Bob to calculate an erroneous level of security. In this case, a notional adversarial party Eve may have more knowledge then Alice and Bob's calculations suggest. Shot-noise is solely present when there is only vacuum in the signal port of the 90$^\circ$ optical hybrid detector with LO in the other port. Due to dynamic system parameters such as fluctuations in LO power, it is necessary that the shot-noise be determined at high-frequency, preferably in real-time. We are able to make secondary use of our clone LOs by making a shot-noise measurement on the clone LO bin 500 ns before or after Alice's signal pulse, see Fig. \ref{cloneLO}. This temporal separation along with orthogonal reference and signal polarizations ensure that no photons from the reference pulse contaminate the shot-noise estimation. As seen in Fig. \ref{shot-noise} \footnote{The actual conversion from the mV input to the ADC and the integer value reported by the ADC is not required, since we normalize to this shot-noise value.}, we generate the expected linear relationship between the LO power and the shot-noise. This data was taken while blocking the signal input port on the $90^\circ$ optical hybrid. We varied the True LO power and monitored it with a power meter from an optical tap while recording the variance at each power with the $90^\circ$ optical hybrid. The electrical noise was measured with both $90^\circ$ optical hybrid ports blocked. The final shot-noise reported is the measured variance minus the electrical noise. A more secure version of this method would be to introduce an optical switch at Bob's input to eliminate the possibility of Eve tampering with the shot-noise measurement. 

\subsection{Polarization Correction}
When the polarization of the reference light from Alice matches the polarization of Bob's reference LO the Z values are maximized. Thus, the Z value itself provides a metric we can use to correct the polarization. The polarization controller consists of 4 piezoelectric actuators, fiber squeezers, which are controlled digitally by the FPGA. An FPGA-based maximization algorithm steps each actuator up or down in voltage in sequence while monitoring changes in Bob's average measured Z values. Steps that result in Z values greater than or equal to the previous Z value result in acceptance of new actuator setting. When Bob is not locked onto Alice's signal, bias removal is imperfect and erroneous triggers may be present. Thus, in addition to monitoring the Z values, we count the number of reference pulses measured within a fixed interval which has a known value for 1 MHz operation. The algorithm only accepts actuator settings which bring the cycle number closer to or equal to the correct value.
\begin{figure}[t]
\includegraphics[width=\linewidth]{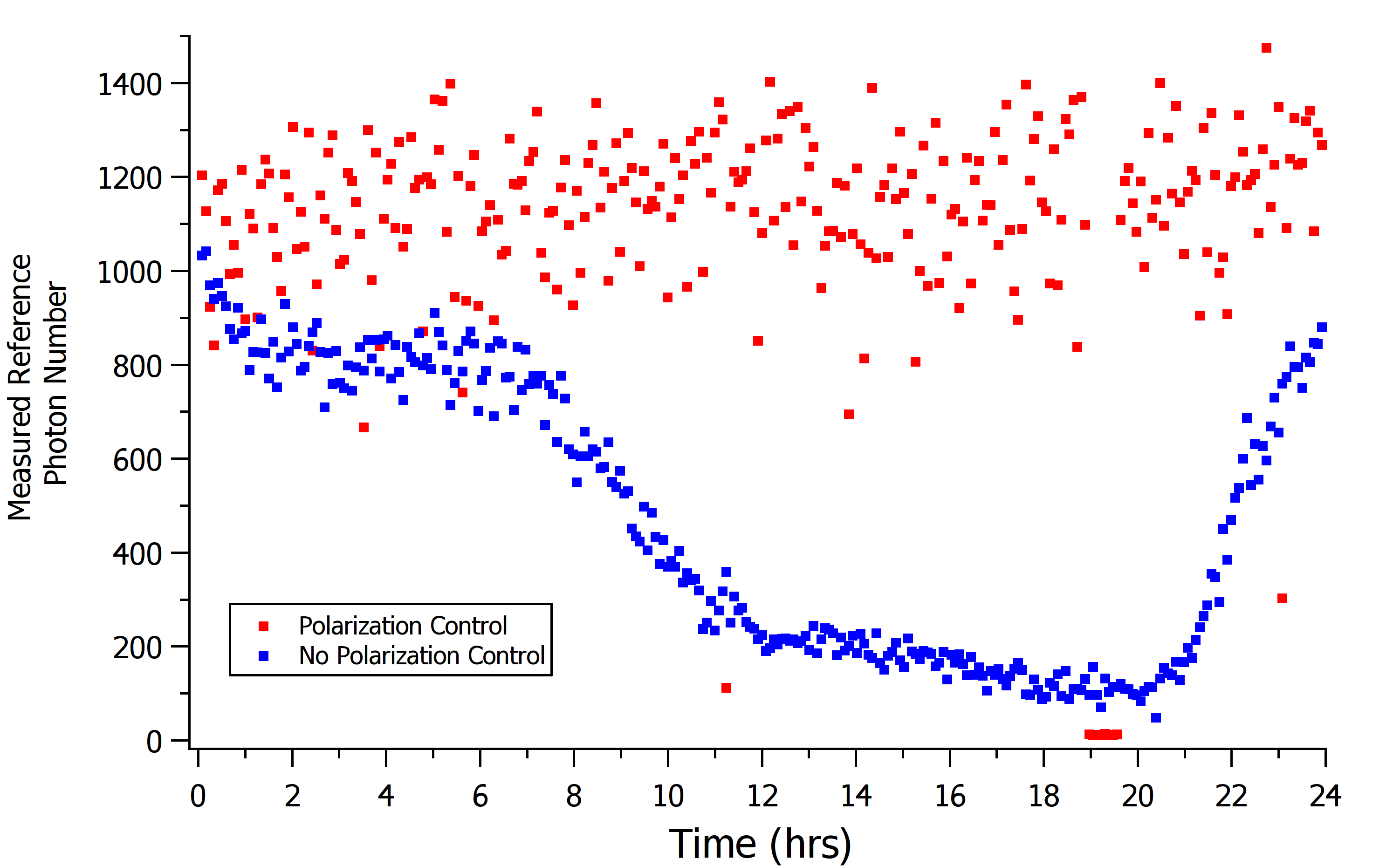}
\caption{Mean reference photon number versus time in the case of active and inactive polarization stabilization. }
\label{figPol}
\end{figure}

We performed a 24 hr polarization test in which the reference pulse photon number was measured every 5 minutes. This test was performed over the out-and-back optical fiber path shown in Fig. \ref{figPath}. The test was performed with and without polarization correction. As can be seen in Fig. \ref{figPol}, with no polarization correction the polarization is relatively stable between successive measurements at 5 minute intervals. However, over a 24 hr period the temperature change does cause polarization dependent loss in our measured reference photon number, the reference pulse polarization no longer matches Bob's reference LO polarization which is fixed. With polarization correction active, we do get successful polarization correction that counteracts the changes due to temperature. As seen in Fig. \ref{figPol}, the polarization correction algorithm leads to a larger variation in the measured reference photon number compared with no polarization correction. Due to this increased noise, we turn polarization correction off during CVQKD packet transmission and turn correction on between packets. Additionally, at 19 hrs it appears to lose the reference pulse lock completely and it takes several minutes to regain the lock. Thus, future research efforts will seek improvements to the polarization correction algorithm such as noise reduction, its ability to maintain the reference pulse lock, and its ability to quickly recover from a loss of lock.

\subsection{CV-QKD Packet Structure and Phase Offset ($\delta$) Determination\label{packetStructure}}
The system operates at a static 1 MHz rate in which one reference and one signal pulse are sent from Alice to Bob once every microsecond. This happens regardless of whether Alice and Bob are performing CV-QKD. This continual operation allows the polarization correction and time synchronization algorithms to maintain the link. To indicate the beginning of a CV-QKD packet Alice sends a bright reference pulse, at least twice as bright as a normal reference pulse, that can easily be distinguished from the standard reference pulse. The bright pulse indicates the beginning of a CV-QKD packet. Each packet of signal pulses from Alice contains a header, a timestamp or packet ID, the encoded CV-QKD values, and a footer. The packet header and footer contain a 64-bit binary pattern which is known by Bob and is used to sort his measured bits and identify the 0's from the 1's. This pattern matching allows determination of the unknown phase $\delta$ from Section \ref{Local LO CV-QKD}, since identifying the 0 and 1 clusters enables correction with a rotation. We note that the length of this pattern and the photon number will affect the accuracy of the $\delta$ estimate. We utilized a 64-bit pattern encoded in phase with 100 photons per pulse. The optimal pattern length and photon number are areas of future investigation. If the pattern correlation in both the header and footer is not above a predefined threshold set by Alice and Bob, it is thrown away. Causes for this could be triggering on an erroneous pulse, a sudden loss of timing lock, or probabilistic failure in the case of a small signal photon number. Ideally, the size of the header, ID, and footer relative to the encoded CV-QKD values would be minimized.

\section{Discussion}
We have reported a proof-of-principal demonstration of a true LO CV-QKD system transmitting classical and quantum signals in coexistence over a deployed aerial optical fiber link in which Alice and Bob are separated by 5.2 km. To do so, we solved practical challenges such as detector bias correction, timing synchronization, polarization correction, vacuum noise measurement, and signal-reference relative phase correction. The results of our tests indicate the system can perform the GMCS CV-QKD protocol over distances of up to 40 km.

Improvements to the system which we have already identified include: improving the current system's repetition rate, introducing a low-loss optical switch to isolate Bob's system from the network fiber connection during vacuum measurement, and making Alice's encoding scheme passive \cite{Qi2020}.

\section*{Acknowledgements}
Funding for this work was provided by the U.S. Department of Energy, Office of Cybersecurity Energy Security and Emergency Response (CESER) through the Risk Management Tools and Technologies (RMT) Program. This manuscript has been authored by UT-Battelle, LLC under Contract No. DE-AC05-00OR22725 with the U.S. Department of Energy. The United States Government retains and the publisher, by accepting the article for publication, acknowledges that the United States Government retains a non-exclusive, paid-up, irrevocable, worldwide license to publish or reproduce the published form of this manuscript, or allow others to do so, for United States Government purposes. The Department of Energy will provide public access to these results of federally sponsored research in accordance with the DOE Public Access Plan (http://energy.gov/downloads/doe-public-access-plan).

\appendix

\section{Secret Key Rate with Infinite Signals\label{secretKey}}
We calculate our key-rate using methods given in \cite{lodewyck2007quantum,fossier2009improvement}.
The key-rate for an infinite key-length scenario is
\begin{equation}
R_{\infty}= \beta I_{AB}-\chi_{BE}\textrm{.}
\label{R}\end{equation}
where $\beta$ is the efficiency of the reconciliation protocol,  $I_{AB}$ is Alice and Bob's mutual information, and $\chi_{BE}$ is the Holevo information.

The mutual information for the heterodyne case is
\begin{equation}I^{\textrm{\textrm{het}}}_{AB}=\log_2\frac{V+\chi_{\textrm{tot}}}{1+\chi_{\textrm{tot}}}
\label{mi}\end{equation}
with
\begin{equation}\chi_{\textrm{tot}}=\chi_{\textrm{line}}+\frac{\chi_{\textrm{het}}}{T}\textrm{,}\end{equation}

\begin{equation}\chi_{\textrm{het}}=\frac{1+\left(1-\eta\right)+2 ~\upsilon_\textrm{el}}{\eta}\textrm{,}\end{equation}

\begin{equation}\chi_{\textrm{line}}=\frac{1+T\xi}{T}-1\end{equation}
where $V=V_A+1$, $V_A$ is Alice's modulation variance, $T$ is the channel transmission, $\eta$ is Bob's detection efficiency, $\xi$ is the excess noise before transmission losses, and $\upsilon_\textrm{el}$ is detector noise.

To determine the Holevo information 
\begin{equation}\chi_{BE}=S_E-S_{E|B}\label{holevo1}\end{equation}
we need to calculate entropies associated with the state accessible to Eve for collective measurement $S_E$ and also $S_{E|B}$ which is the entropy of the state after Bob's measurement. When the relevant covariance matrices are known these entropies are of the form
\begin{equation}\chi_{BE}=\sum_{i=1}^2 g\left(\frac{\lambda_i-1}{2}\right)-\sum_{i=3}^5 g\left(\frac{\lambda_i-1}{2}\right)\end{equation}
where
\begin{equation}g\left(x\right)=\left(x+1\right)\log_2\left(x+1\right)-x\log_2\left(x\right)\textrm{,}\end{equation}
\begin{equation}\lambda_{1,2}^2=\frac{1}{2}\left(A\pm\sqrt{A^2-4B}\right)\textrm{,}\end{equation}
\begin{equation}A=V^2\left(1-2T\right)+2T+T^2\left(V+\chi_{\textrm{line}}\right)^2\textrm{,}\end{equation}
\begin{equation}B=T^2\left(V\chi_{\textrm{line}}+1\right)^2\textrm{,}\end{equation}
\begin{equation}\lambda_{3,4}^2=\frac{1}{2}\left(C\pm\sqrt{C^2-4D}\right)\textrm{,}\end{equation}
\begin{align}C=&\frac{1}{T^2 \left(V+\chi_{\textrm{tot}}\right)^2}\times\nonumber\\
&\left[A\chi_{A\textrm{het}}^2+B+1+2T\left(V^2-1\right)\right.\nonumber\\
&\left.\qquad+2\chi_{\textrm{het}}\left(V\sqrt{B}+T\left(V+\chi_{\textrm{line}}\right)\right)\right]\textrm{,}\end{align}
\begin{equation}D=\left(\frac{V+\sqrt{B}\chi_{\textrm{het}}}{T\left(V+\chi_{\textrm{tot}}\right)}\right)^2\textrm{,}\end{equation}
and $\lambda_5 = 1$.

\section{Secret Key Rate with Finite Signals\label{secretKeyFS}}
 A finite-size of signal exchanges are used by Alice and Bob to estimate  the channel transmission and the excess noise. Thus, there is also uncertainty in these estimations. Security dictates that we choose a conservative estimate of these quantities, which reduces the key rate. With more signal exchanges, more data, the error bars tighten which improves the key-rate. We apply the finite-size analysis detailed in \cite{leverrier2010finite}.

The key-rate for a finite-size scenario is
\begin{equation}
R=\frac{n}{N}\left[ \beta I_{AB}-\chi_{BE}^{\textrm{max}}\left(m\right)-\Delta(n)\right]\label{fs_rate}
\end{equation}
where $\beta$ is the efficiency of the reconciliation protocol, $m$ is the number of pulses used for parameter estimation, $n$ is the number of pulses available for raw key, $N=m+n$ is the total pulse number,  $I_{AB}$ is Alice and Bob's mutual information, $\chi_{BE}\left(m\right)$ is the maximum Holevo information associated with parameter estimation using $m$ pulses, and $\Delta(n)$ is related to the security of the privacy amplification given raw key $n$ and is approximately
\begin{equation}\Delta(n)\approx7\sqrt{\frac{\log_2\left(2/\bar{\epsilon}\right)}{n}}\end{equation}
with $\bar{\epsilon}=10^{-10}$. In principle, using an infinite number of signal exchanges allows the fraction used for parameter estimation to approach zero and $n/N\rightarrow 1$. In which case the rate would go to the familiar infinite key-rate 
\begin{equation}
R_{\infty}= \beta I_{AB}-\chi_{BE}\textrm{.}
\end{equation}
However, for practical application we  assume half the key is used for parameter estimation. Thus with increasing number of pulses we asymptotically approach the rate  
\begin{equation}R_{1/2}^{\textrm{max}}=\frac{1}{2}\left[ \beta I_{AB}-\chi_{BE}\right]\textrm{.}
\end{equation}

The method of Leverrier et al. \cite{leverrier2010finite} determines the minimum \emph{true} value of the effective transmission \begin{equation}T_{min}=\left(\sqrt{\hat{T}}-z \sqrt{\frac{\hat{\sigma}^2}{m V_A}}\right)^2\label{min_t}\end{equation}
and the maximum \emph{true} value of the noise 
\begin{equation}\sigma^2_{max}=\hat{\sigma}^2+z\frac{\hat{\sigma}^2\sqrt{2}}{\sqrt{m}}\label{max_noise}\end{equation} 
given that we estimate expected values $\hat{T}$ and $\hat{\sigma^2}$, we use $m$ data points for estimation, $z=6.5$ (see \cite{leverrier2010finite}), and that these are not the minimum or maximum values, respectively, with probability $\epsilon_{PE}/2$ with $\epsilon_{PE}=10^{-10}$. Again, $\hat{T}$ and $\hat{\sigma}^2$ are the most-likely estimates given our measurement result, and the true transmission is greater than $T_{min}$ and the true noise is less than $\sigma^2_{max}$ with probability $1-\epsilon_{PE}/2$.

The expected effective transmission value determined by Alice and Bob is
\begin{equation}\hat{T}=\frac{\eta T}{2}\label{exp_t}\end{equation}
with detector efficiency $\eta$, channel transmission $T$, and factor of $1/2$ due to heterodyne detection.

The expected noise is
\begin{equation}\hat{\sigma}^2=1+T\xi=1+\frac{\eta T}{2}\xi+\xi_d\label{exp_noise}\end{equation}
with the measured detector noise $\xi_d$ and an excess noise that is assumed to be phase noise given by
\begin{equation}\xi=\xi_{p}= V_A \Delta\phi\textrm{.}\end{equation}

The expected values in Eq. (\ref{exp_t}) and Eq. (\ref{exp_noise}) are used in the mutual information as before. However, to calculate the Holveo bound we apply our conservative finite-size estimates from (\ref{min_t}) and (\ref{max_noise}). Our rate is then given by Eq. (\ref{fs_rate}). 

\section{Gaussian Random Number Generation\label{secretKeyFS}}

Gaussian random variables are generated using the inversion principle \cite{Devr86}. The inversion principle allows us to generate a Rayleigh distributed random number
\begin{equation}r=\sigma\sqrt{-\ln u}\label{generateR}\end{equation}
where $u$ is a uniformly distributed random number and $\sigma$ is a static scaling parameter. The Rayleigh distribution is given as \begin{equation}f\left(r\right)=\frac{r}{\sigma^2}e^{-\frac{r^2}{2\sigma^2}}\textrm{.}\label{Rayleigh}\end{equation}

Let the Rayleigh distributed value be the radial value of our $x,p$ coordinates, $r=\sqrt{x^2+p^2}$. Next, we generate a second uniform value $\theta\in\left\{0,2\pi\right\}$ with which we fully define our coordinates
\begin{align}
x&=r \cos\theta\\
p&=r \sin\theta\textrm{.}
\end{align}

The probability distribution for r is the Rayleigh distribution given in Eq. \ref{Rayleigh} and the distribution for $\theta$ is uniform 
\begin{equation}
\Theta\left(\theta\right)=\frac{1}{2\pi}\textrm{.}
\end{equation}

The distribution for these combined is just their product
\begin{equation}F\left(r,\theta\right)=f\left(r\right)\times\Theta\left(\theta\right)=\frac{r}{2\pi\sigma^2}e^{-\frac{r^2}{2\sigma^2}}\end{equation}
which under coordinate transform $r=\sqrt{x^2+p^2}$ and $\theta=\arctan\left(\frac{p}{x}\right)$ with Jacobian $1/r$ 
gives this distribution in terms of $x,p$ as
\begin{equation}F\left(x,p\right)=\frac{1}{2\pi\sigma^2}e^{-\frac{x^2+p^2}{2\sigma^2}}=g\left(x\right)\times g\left(p\right)\end{equation}
with Gaussian distribution
\begin{equation}
g\left(q\right)=\frac{1}{\sqrt{2\pi\sigma^2}}e^{-\frac{q^2}{2\sigma^2}}\label{Gaussian}\end{equation}
as required.

Specific to our experiment, the uniform values to generate $r$ and $\theta$ are each 16-bit. 

\section{Bias Free Phase-Amplitude Modulator\label{phase-amplitude}}
To derive the modulation performed by the bias-free phase-amplitude modulator we reference single-mode coherent state relations given in \cite{loudon2000quantum}. We assume an incident coherent state in a single-mode.  
Referencing Fig. \ref{amplitudePhase}, the relevant annihilation operator relationship for light incident into port 1 or 2 is
\begin{equation}
\begin{bmatrix}
\hat{a}_3 \\
\hat{a}_4 
\end{bmatrix} = \textrm{BS}\cdot\begin{bmatrix}
\hat{a}_1 \\
\hat{a}_2 
\end{bmatrix}
\end{equation}
where
\begin{equation}
\textrm{BS}=\frac{1}{\sqrt{2}}\begin{bmatrix}
1 & i \\
i & 1 
\end{bmatrix}
\textrm{.}\end{equation}
Due to the temporal delay, 100 ns, between the CCW and CW propagating pulses different phases can be applied to each, see \cite{qi2018noise}. We can then evolve our annihilation operators further with these phase modulations as
\begin{equation}
\begin{bmatrix}
\hat{a}_{cw} \\
\hat{a}_{ccw}
\end{bmatrix} = \begin{bmatrix}
e^{i\phi_{cw}} & 0 \\
0 & e^{i\phi_{ccw}} 
\end{bmatrix}\cdot\begin{bmatrix}
\hat{a}_3 \\
\hat{a}_4 
\end{bmatrix}
\end{equation}
Due to the Sagnac configuration light propagating CCW becomes an input at BS port 3 and CW propagating light is incident at port 4. The annihilation operators for light exiting the Sagnac loop are then
\begin{equation}
\begin{bmatrix}
\hat{a}_{1}' \\
\hat{a}_{2}' 
\end{bmatrix} = \textrm{BS}^{\dagger}\cdot\begin{bmatrix}
\hat{a}_{ccw} \\
\hat{a}_{cw} 
\end{bmatrix}
\end{equation}
Carrying out all these operations the annihilation operators exiting the Sagnac loop are
\begin{equation}
\begin{bmatrix}
\hat{a}_{1}' \\
\hat{a}_{2}' 
\end{bmatrix} 
= e^{i\gamma_{+}}
\begin{bmatrix}
\left(\hat{a}_{1}\cos\gamma_{-}+\hat{a}_{2}\sin\gamma_{-}\right) \\
\left(-\hat{a}_{1}\sin\gamma_{-}+\hat{a}_{2}\cos\gamma_{-}\right)\end{bmatrix}
\end{equation}
where $\gamma_\pm=\frac{\phi_{\textrm{ccw}}\pm\phi_{\textrm{cw}}}{2}$. We can now solve this system for $\hat{a}_1$ and $\hat{a}_2$.
\begin{equation}
\begin{bmatrix}
\hat{a}_{1} \\
\hat{a}_{2} 
\end{bmatrix} 
= e^{-i\gamma_{+}}
\begin{bmatrix}
\left(\hat{a}_{1}'\cos\gamma_{-}-\hat{a}_{2}'\sin\gamma_{-}\right) \\
\left(\hat{a}_{1}'\sin\gamma_{-}+\hat{a}_{2}'\cos\gamma_{-}\right)\end{bmatrix}
\label{output}\end{equation}

The incident coherent state is
\begin{equation}
\left|\alpha\right\rangle=\hat{D}_1\left(\alpha\right)\left|0\right\rangle=e^{\alpha \hat{a}_1^\dagger -\alpha^{*}\hat{a}_1}\left|0\right\rangle
\end{equation}
with displacement operator $D$. Rewriting the terms inside the displacement operator with Eq. \ref{output} we get
\begin{align}\alpha \hat{a}_1^\dagger -&\alpha^{*}\hat{a}_1=\\
&\alpha e^{i\gamma_{+}}\left({\hat{a}_{1}}^{'\dagger}\cos\gamma-{\hat{a}_{2}}^{'\dagger}\sin\gamma\right)\\
&\quad-\alpha^{*}e^{-i\gamma_{+}}\left( \hat{a}_{1}'\cos\gamma-\hat{a}_{2}'\sin\gamma\right)\nonumber\\
&\cos\gamma \left( \alpha e^{i\gamma_{+}}{\hat{a}_{1}}^{'\dagger}-\alpha^{*}e^{-i\gamma_{+}}{\hat{a}_{1}'}\right)\\
&\quad-\sin\gamma \left( \alpha e^{i\gamma_{+}}{\hat{a}_{2}}^{'\dagger}-\alpha^{*}e^{-i\gamma_{+}}{\hat{a}_{2}'}\right)
\end{align}
We then can then write the input displacement operator in terms of the outputs as
\begin{align}
&\hat{D}_1\left(\alpha\right)=\hat{D}_{1'}\left(\cos\gamma \alpha e^{i\gamma_{+}}\right)\times\hat{D_{2'}}\left(-\sin\gamma \alpha e^{i\gamma_{+}}\right)\textrm{.}\end{align}
As seen above, the coherent state at port 2' is represented by complex value 
\begin{align}\alpha_{2'}&=  e^{i\gamma_{+}}\sin\left(-\gamma\right) \alpha\nonumber\\
&=e^{i\frac{\left(\phi_{\textrm{cw}} + \phi_{\textrm{ccw}}\right)}{2}}\sin\left(\frac{\phi_{\textrm{cw}}-\phi_{\textrm{ccw}}}{2}\right)\alpha\textrm{.}\end{align}

\bibliographystyle{apalike}
\bibliography{main.bib}
\end{document}